\documentclass[aps,prb,superscriptaddress,reprint,showpacs,floatfix]{revtex4-2}
\usepackage{graphicx}
\usepackage{dcolumn}
\usepackage{bm}
\usepackage{xcolor}
\usepackage{relsize}
\usepackage{amsmath}
\usepackage{amsfonts}
\usepackage{amssymb}
\usepackage{mathtools}
\usepackage{braket}
\usepackage{gensymb}
\usepackage{url}
\usepackage{mathrsfs}
\usepackage{enumitem}
\setlist[itemize]{label={}}

\usepackage[colorlinks = true,
            linkcolor = blue,
            urlcolor  = blue,
            citecolor = blue,
            anchorcolor = blue]{hyperref}
\usepackage{footmisc}
\usepackage[utf8]{inputenc}

\begin{document}

\preprint{APS/123-QED}

\title{Spin relaxation in a polariton fluid: quantum hydrodynamic approach}

\author{D. A. Saltykova}
\affiliation{ITMO University, St. Petersburg 197101, Russia}

\author{A. V. Yulin}
\affiliation{ITMO University, St. Petersburg 197101, Russia}

\author{I. A. Shelykh}

\affiliation{Science Institute, University of Iceland, Dunhagi 3, IS-107 Reykjavik, Iceland}

\date{\today}

\begin{abstract}

Cavity polaritons, the elementary excitations appearing in quantum microcavities in the strong-coupling regime, reveal clear signatures of quantum collective behavior. The combination of unique spin structure and strong nonlinear response opens the possibility of direct experimental observation of a plethora of nontrivial optical polarization phenomena. Spin relaxation processes are of crucial importance here. However, a mathematical formalism for their coherent description is still absent. In the present paper, based on the quantum hydrodynamics approach for a two-component liquid, we derive the set of the corresponding equations where both energy and spin relaxation terms appear naturally.  We analyze in detail how these terms affect the dynamics of spinor polariton droplets in the external magnetic field and the dispersion of elementary excitations of a uniform polariton condensate. Although we focus on the case of cavity polaritons, our approach can be applied to other cases of spinor bosonic condensates, where the processes of spin relaxation play a major role.
\end{abstract}

\maketitle

\section{Introduction}

The physics of quantum fluids is among the major trends of modern condensed matter and atomic physics. The major step forward here was done back in the 90s when experimental achievement of an atomic Bose-Einstein condensate was reported. However, the characteristic temperatures of the corresponding phase transition for cold atomic gases fall into the nanokelvin range, and, although its laboratory realization does not currently represent a major problem, related application perspectives are still challenging. This motivated the search for analogues of BECs in other material platforms, optical or condensed matter, where effective masses of bosonic elementary excitations are several orders of magnitude smaller, and one can expect formation of quantum collective states up to the temperatures.

In this context, an attractive possibility is presented by planar quantum microcavities, specifically designed structures where the interaction between a spatially confined optical mode with excitons is dramatically enhanced. This allows the achievement of the regime of strong light-matter coupling, where exciton polaritons (also known as cavity polaritons), hybrid half-light half-matter particles, appear as elementary excitations \cite{Kavokin2017_OxfPr,Carusotto2013}. They obtain an extremely small effective mass (about $10^{-5}$ of the mass of free electrons) and a long coherence length (on the mm scale) \cite{Ballarini2017} from the cavity photons, while the presence of an excitonic component makes possible the polariton-polariton interactions responsible for the robust nonlinear optical response. The polariton BEC and superfluidity was experimentally observed at remarkably high temperatures under both optical \cite{Kasprzak2006,Balili2007,Christopoulos2007} and electrical \cite{Schneider2013} excitation.

Compared with cold atoms, polaritons reveal the following important properties. 

First, they can be directly created optically and have finite lifetimes because of the possibility for the photons to leave the system through the partially transparent Bragg mirrors. This means that all major phenomena in microcavities are necessarily of dynamic nature and are governed by a delicate interplay of external pumping, radiative decay, and other types of damping. 

Second, the presence of the excitonic fraction in a polariton makes possible efficient polariton-phonon interaction, which can both couple polaritons to an ensemble of incoherent excitons and lead to the energy relaxation within the polariton liquid itself. 

Third, polaritons possess a peculiar spin (or rather pseudo-spin) structure \cite{ShelykhReview}. Similarly to photons, polaritons have two possible spin projections on the structure growth axis corresponding to the two opposite circular polarizations, which can be mixed by effective magnetic fields of various origin. 
The real magnetic field applied along the structure growth axis and acting on the excitonic component splits the polariton states in energy with opposite circular polarizations (Zeeman splitting), while the TE-TM splitting of the photonic modes of a planar resonator couples these states to each other via a $k$-dependent term, thus playing the role of an effective spin-orbit interaction \cite{ShelykhReview}. The presence of effective magnetic fields can lead to both polariton spin precession in the conservative regime and spin alignment if spin relaxation processes play a role. 

Quite importantly, polariton-polariton interactions are spin-dependent as well. Indeed, they stem from the interactions of their excitonic components, which are dominated by the exchange term \cite{Ciuti1998}.
This leads to the fact that polaritons of the same circular polarization interact orders of magnitude stronger than polaritons with opposite circular polarizations \cite{Glazov2009}. This leads, in particular, to the spontaneous appearance of linear polarization in polariton BECs \cite{Baumberg2008}, the so-called self-induced Larmor precession of polarization in elliptically pumped polariton condensates \cite{ShelykhSILP} and full paramagnetic screening, also known as the spin Meissner effect, in polariton condensates in the external magnetic field \cite{PLARubo,Larionov2010,Walker2011,Fischer2014}.

Naturally, any sound description of polarization-related phenomena in microcavities requires a correct account of the processes of spin relaxation within polariton condensates. However, although the spin structure of the polaritons and the presence of the corresponding effective magnetic fields can be directly accounted for in Gross-Pitaevskii type of description via replacement of the scalar macroscopic wavefunction by a spinor \cite{Shelykh2006,Borgh2010}, the introduction of pure spin relaxation is still puzzling. The present paper represents an attempt to construct the corresponding theory.

Similarly to our previous work on pure energy relaxation in the scalar case, we base our consideration on a quantum hydrodynamics approach, describing the systems dynamics via a set of classical field Hamilton equations for canonically conjugated variables of concentrations and phases of two spin components and introducing relaxation in a natural way by adding the gradient terms. We analyze how these terms affect the dynamics of the polariton droplets in the magnetic field and the dispersion of elementary excitations.  

As our goal here is to focus on the role of pure spin relaxation only, and for the reason of clarity of the presentation, we will neglect all other dissipative processes, such as finite lifetimes, external pumping, and coupling with an incoherent excitonic reservoir. They can be easily incorporated into the model if necessary.

\section{Derivation of the dynamic equations}



We consider a generalized model of a conservative spinor Bose–Einstein condensate (BEC) of polaritons, which is described by the following system of coupled Gross–Pitaevskii equations:

\begin{widetext}
\begin{equation} \label{GPEq}
     \begin{cases}
    i \partial_t \psi_1= - \frac{1}{2} (\partial_x^2+\partial_y^2)\psi_1 + \alpha_1|\psi_1|^2\psi_1+\alpha_2|\psi_2|^2\psi_1-\frac{\Delta}{2}\psi_1-\delta \psi_2, \\
    
    i \partial_t \psi_2= - \frac{1}{2} (\partial_x^2+\partial_y^2)\psi_2 +\alpha_1|\psi_2|^2\psi_2+\alpha_2|\psi_1|^2\psi_2+\frac{\Delta}{2}\psi_2-\delta \psi_1. 
    \end{cases}
\end{equation}
\end{widetext}

Here, $\psi_1$ and $\psi_2$ denote the condensate wave functions corresponding to the two circular polarization components. The first term on the right-hand side of each equation represents the standard kinetic energy operator in two spatial dimensions. The nonlinear interaction terms include both self-phase modulation $\alpha_1$ and cross-phase modulation $\alpha_2$, where $\alpha_1$ and $\alpha_2$ characterize the strengths of intra-component and inter-component interactions, respectively. The parameter $\Delta$ denotes the Zeeman splitting, which introduces an energy offset between the spin components due to the presence of an external magnetic field. The coupling constant $\delta$ accounts for the linear polarization splitting in the microcavity, which breaks the rotational symmetry in the $xy$-plane and leads to anisotropic behaviour in the condensate dynamics.

Let us pay attention to the fact that the coupled Gross–Pitaevskii equations for the conservative spinor case (see Eq.~\ref{GPEq}) represent an equation for a classical field. These equations can be derived from the principle of least action,
$
\delta S = 0, \quad S = \int \mathscr{L} \, d^2\mathbf{r} \, dt,$
where \( \mathscr{L} \) is the corresponding Lagrangian density:
\begin{widetext}
\begin{equation} \label{Lagrangian}
\begin{split}
   \mathscr{L}= \frac{i}{2}(\psi_1^{*}\frac{\partial \psi_1}{\partial t} - \psi_1 \frac{\partial \psi_1^{*}}{\partial t})+  \frac{i}{2}(\psi_2^{*}\frac{\partial \psi_2}{\partial t} - \psi_2 \frac{\partial \psi_2^{*}}{\partial t})-\frac{1}{2} (| \nabla \psi_1|^2+|\nabla \psi_2|^2)- \\ -\frac{\alpha_1}{2}(|\psi_1|^4+|\psi_2|^4)-\alpha_2|\psi_2\psi_1|^2 +\frac{\Delta}{2}(|\psi_1|^2-|\psi_2|^2)+\delta(\psi_1 \psi_2^\ast+\psi_2 \psi_1^\ast), 
\end{split}
\end{equation}
\end{widetext}
In our notation $| \nabla \psi_{1,2}|^2=|\partial_x \psi_{1,2}|^2+|\partial_y \psi_{1,2}|^2$.

By introducing the Madelung representation, where each field is expressed in terms of its amplitude and phase, both depending on the spatial coordinates \mbox{$\vec{r}=(x,y)$} and time:
\begin{equation}
\begin{cases}
    \psi_1(\vec{r},t) = \sqrt{\rho_1(\vec{r},t)} e^{-i \theta_1(\vec{r},t)}, \\
     \psi_2(\vec{r},t) = \sqrt{\rho_2(\vec{r},t)} e^{-i \theta_2(\vec{r},t)},
     \end{cases}
\end{equation}

we obtain for the Lagrangian:
\begin{widetext}
\begin{equation} \label{LagrMadel}
\begin{split}
\mathcal{L} =\ 
  \rho_1 \partial_t \theta_1 +\rho_2 \partial_t \theta_2  - \frac{1}{2} \left[ \rho_1 (\nabla \theta_1)^2 + \rho_2 (\nabla \theta_2)^2 \right] 
- \frac{1}{2} \left[ \frac{(\nabla \rho_1)^2}{4\rho_1} + \frac{(\nabla \rho_2)^2}{4\rho_2} \right]  -\\-\frac{\alpha_1}{2} (\rho_1^2 + \rho_2^2) - \alpha_2 \rho_1 \rho_2 + \frac{\Delta}{2} (\rho_1 - \rho_2) +2 \delta \sqrt{\rho_1\rho_2} \cos(\theta_1-\theta_2).
\end{split}
\end{equation}
\end{widetext}

From this expression it follows that the phases $\theta_{1,2}$ can be treated as generalized coordinates of the field, while the densities $\rho_{1,2}$ correspond to the canonically conjugate momenta, given by
\begin{equation}
\begin{cases}
    \pi_1=\frac{\partial\mathscr{L}}{\partial(\partial_t\theta_1)}=\rho_1, \\
    \pi_2=\frac{\partial\mathscr{L}}{\partial(\partial_t\theta_2)}=\rho_2.
     \end{cases}
\end{equation}

To simplify the analysis and gain physical insight, it is convenient to introduce a change of variables that separates the collective and internal degrees of freedom in the two-component condensate. Specifically, we define a new set of variables in terms of the original density and phase fields: 
\begin{equation}
\begin{cases} \label{centerofmass}
     \theta=\theta_1-\theta_2, \\
     \Theta=\frac{1}{2}(\theta_1+\theta_2), \\
    \Pi=\rho_1+\rho_2, \\
     \mathcal{Z}=\frac{1}{2}(\rho_1-\rho_2). \\
\end{cases}
\end{equation}
Here, \( \Theta \) denotes the global phase of the condensate and describes the collective dynamics. Variable \( \theta \) represents the relative phase between the two components and characterizes internal coherence and interference phenomena. The quantity \( \Pi \) corresponds to the total density of the condensate, determining its macroscopic hydrodynamic behavior. In contrast, \( \mathcal{Z} \) quantifies the density imbalance between components and effectively plays the role of pseudospin polarization. Mathematically, the transformation \ref{centerofmass} is analogous to introducing the center of mass and relative coordinates in classical mechanics. The Lagrangian \ref{LagrMadel} can be reformulated in terms of the new variables as follows:
\begin{equation}
\begin{aligned}
&\mathcal{L} =  \, \Pi \partial_t \Theta + \mathcal{Z} \partial_t \theta - \frac{1}{2} [\Pi (\nabla \Theta)^2 +\Pi (\frac{\nabla \theta}{2})^2] - \mathcal{Z} \nabla \Theta \nabla \theta  -\\ &-\frac{1}{8} \left[
\frac{ \left( \frac{1}{2} \nabla \Pi + \nabla \mathcal{Z} \right)^2 }{ \frac{1}{2} \Pi + \mathcal{Z} }
+
\frac{ \left( \frac{1}{2} \nabla \Pi - \nabla \mathcal{Z} \right)^2 }{ \frac{1}{2} \Pi - \mathcal{Z} }
\right] -\frac{\alpha_1+\alpha_2}{4} \Pi^2-
\\&-(\alpha_1 -\alpha_2)\mathcal{Z}^2  + \Delta \mathcal{Z}+\delta\sqrt{\Pi^2-4\mathcal{Z}^2}\cos \theta.
\end{aligned}
\end{equation}

In accordance with the canonical formalism, the global phase $\Theta$ and the relative phase $\theta$ are regarded as generalized coordinates, and their corresponding canonically conjugate momenta are:
\begin{equation}
\begin{cases}
    \pi_\theta=\frac{\partial\mathscr{L}}{\partial(\partial_t\theta)}=\mathcal{Z}, \\
    \pi_\Theta=\frac{\partial\mathscr{L}}{\partial(\partial_t\Theta)}=\Pi.
     \end{cases}
\end{equation}

Hence, the Hamiltonian density, expressed in terms of the transformed fields $(\theta, \Theta, \Pi, \mathcal{Z})$, takes the form:
\begin{equation}
\begin{aligned}
\mathscr{H} &= \pi_\theta \partial_t \theta + \pi_\Theta \partial_t \Theta - \mathscr{L} =\\
=& \frac{1}{2} \left[ \Pi (\nabla \Theta)^2 + \Pi \left( \frac{\nabla \theta}{2} \right)^2 \right] 
    + \mathcal{Z} \nabla \Theta \nabla \theta + \\
    \quad +& \frac{1}{8} \left[ \frac{\left( \frac{1}{2} \nabla \Pi + \nabla \mathcal{Z} \right)^2}{\frac{1}{2} \Pi + \mathcal{Z}} + \frac{\left( \frac{1}{2} \nabla \Pi - \nabla \mathcal{Z} \right)^2}{\frac{1}{2} \Pi - \mathcal{Z}} \right]  +\frac{\alpha_1 + \alpha_2}{4} \Pi^2 + \\+&(\alpha_1 - \alpha_2) \mathcal{Z}^2 - \Delta \mathcal{Z}-\delta\sqrt{\Pi^2-4\mathcal{Z}^2}\cos \theta.
\end{aligned}
\end{equation}

The dynamic field equation can be represented in the Hamiltonian form as

\begin{align}
\begin{cases}    &\partial_t \pi_\Theta=\partial_t \Pi=-\frac{\delta\mathscr{H}}{\delta\Theta},\\
\\
&\partial_t \pi_\theta=\partial_t \mathcal{Z}=-\frac{\delta\mathscr{H}}{\delta\theta},\\ 
\\
&\partial_t \Theta=\frac{\delta\mathscr{H}}{\delta\pi_\Theta}=\frac{\delta\mathscr{H}}{\delta\Pi},\\
\\
&\partial_t \theta=\frac{\delta\mathscr{H}}{\delta\pi_\theta}=\frac{\delta\mathscr{H}}{\delta\mathcal{Z}}.
\end{cases}
\end{align}

After calculating the corresponding functional derivatives for each equation, we obtain:
\begin{widetext}
\begin{align}
\begin{cases}    &\partial_t \Pi=\nabla(\Pi\nabla \Theta)+\nabla(\mathcal{Z} \nabla \theta),\\
\\
&\partial_t \mathcal{Z}=\frac{1}{4}\nabla(\Pi\nabla \theta)+\nabla(\mathcal{Z} \nabla \Theta)-\delta \sqrt{\Pi^2-4 \mathcal{Z}^2}\sin \theta ,\\
\\
&\partial_t \Theta= \frac{1}{2} \left[ (\nabla \Theta)^2 +  \left( \frac{\nabla \theta}{2} \right)^2 \right] - \frac{1}{8} \left[ \frac{1}{2} \frac{\left( \frac{1}{2} \nabla \Pi + \nabla \mathcal{Z} \right)^2}{(\frac{1}{2} \Pi + \mathcal{Z})^2} + \frac{1}{2} \frac{\left( \frac{1}{2} \nabla \Pi - \nabla \mathcal{Z} \right)^2}{(\frac{1}{2} \Pi - \mathcal{Z})^2} \right] + \frac{\alpha_1+\alpha_2}{2}{\Pi} - \frac{1}{8} \nabla  \left[
\frac{ \frac{1}{2} \nabla \Pi + \nabla \mathcal{Z}  }{ \frac{1}{2} \Pi + \mathcal{Z} }
+
\frac{ \frac{1}{2} \nabla \Pi - \nabla \mathcal{Z}  }{ \frac{1}{2} \Pi - \mathcal{Z} }
\right] -\frac{\delta \Pi}{\sqrt{\Pi^2-4\mathcal{Z}^2}}  \cos \theta,\\ \\
&\partial_t \theta= \nabla \Theta \nabla \theta +2(\alpha_1-\alpha_2)\mathcal{Z}-\Delta - \frac{1}{8} \left[  \frac{\left( \frac{1}{2} \nabla \Pi + \nabla \mathcal{Z} \right)^2}{(\frac{1}{2} \Pi + \mathcal{Z})^2} - \frac{\left( \frac{1}{2} \nabla \Pi - \nabla \mathcal{Z} \right)^2}{(\frac{1}{2} \Pi - \mathcal{Z})^2} \right] -\frac{1}{8} \nabla  \left[
2 \frac{ \frac{1}{2} \nabla \Pi + \nabla \mathcal{Z}  }{ \frac{1}{2} \Pi + \mathcal{Z} }
-2
\frac{ \frac{1}{2} \nabla \Pi - \nabla \mathcal{Z}  }{ \frac{1}{2} \Pi - \mathcal{Z} }
\right]+\frac{4 \delta \mathcal{Z}}{\sqrt{\Pi^2-4\mathcal{Z}^2}} \cos \theta.
\end{cases}
\end{align}
\end{widetext}

To more accurately model the nonequilibrium dynamics of the condensate, it is necessary to incorporate relaxation effects into the equations of motion. Mathematically, relaxation is introduced phenomenologically through gradient-flow terms proportional to the functional derivatives of the Hamiltonian. The corresponding modifications to the equations for $\Theta$ and $\theta$ are readily obtained:
\begin{widetext}
\begin{align*}
\begin{cases}    
&\partial_t \Theta=\frac{\delta\mathscr{H}}{\delta\Pi} - \gamma_\Theta\frac{\delta\mathscr{H}}{\delta \Theta}=\frac{\delta\mathscr{H}}{\delta\Pi} + \gamma_\Theta[\nabla(\Pi\nabla \Theta)+\nabla(\mathcal{Z} \nabla \theta)],\\
\\
&\partial_t \theta=\frac{\delta\mathscr{H}}{\delta\mathcal{Z}}-  \gamma_\theta\frac{\delta\mathscr{H}}{\delta \theta}=\frac{\delta\mathscr{H}}{\delta\mathcal{Z}}+  \gamma_\theta[\frac{1}{4}\nabla(\Pi\nabla \theta)+\nabla(\mathcal{Z} \nabla \Theta)-\delta \sqrt{\Pi^2-4 \mathcal{Z}^2}\sin \theta].
\end{cases}
\end{align*}
\end{widetext}

The terms $\gamma_\Theta \frac{\delta \mathscr{H}}{\delta \Theta}$ and $\gamma_\theta \frac{\delta \mathscr{H}}{\delta \theta}$ represent relaxation corrections, reflecting the fact that the system no longer strictly conserves energy but tends to a minimum of energy --- in the direction of the gradient in phases.

Next, we discuss the inclusion of relaxation in the two remaining equations. We will not include relaxation in the equation for the total density $\Pi=\rho_1+\rho_2$, as doing so would contradict the law of conservation of particle numbers in a closed system. However, it is worth including relaxation in the equation for $\mathcal{Z}=\frac{1}{2}(\rho_1-\rho_2)$, because the relative density describes the redistribution of particles between the components of a two-component condensate. This redistribution does not affect the total number of particles, but can occur, for example, as a result of interaction with an external field, intercomponent scattering, or inelastic processes between the components. Such processes are naturally described by relaxation, which tends to equalize densities or bring the system to thermodynamic equilibrium. To incorporate relaxation into the equation for $\mathcal{Z}$, we represent the relative density as:
\begin{equation*}
     \mathcal{Z}=\frac{1}{2}(\rho_1-\rho_2)=\frac{1}{2} \Pi \cos \Omega.
\end{equation*}
 This replacement imposes certain restrictions on the values of relative density, and now $\mathcal{Z} \in \frac{1}{2}[-\Pi;\Pi]$. This is done to exclude a situation where the difference in densities exceeds their sum, and we can get a nonphysical result. To account for dissipative processes, we recast the equation for $\mathcal{Z}$ using the variables $\Pi$ and $\Omega$, from which $\partial_t \Omega$ can be explicitly obtained (see Supplementary Materials). The relaxation terms, which describe the tendency of the system to minimize the Hamiltonian in the direction of change of $\Omega$, are then added:

\begin{align}
     &\partial_t \Omega = \nabla \Omega  \nabla \Theta - \frac{\nabla  \Pi \nabla \theta}{2 \Pi} \sin \Omega -\frac{\nabla^2 \theta}{2} \sin \Omega-\\ \notag
     &-\frac{1}{2} \cos \Omega \nabla \Omega \nabla \theta +2 \delta \sin \theta -\gamma_\Omega \frac{\delta \mathscr{H}}{\delta \Omega}. 
\end{align}

Let us write the Hamiltonian as a function of the fields $\Theta, \ \theta, \ \Pi$ and $\Omega$:
\begin{widetext}
\begin{align}
\mathscr{H} & = \frac{1}{2} \left[ \Pi (\nabla \Theta)^2 + \Pi \left( \frac{\nabla \theta}{2} \right)^2 \right] 
    + \frac{1}{2}\Pi \cos \Omega \nabla \Theta \nabla \theta  + \frac{1}{8} \left[ \frac{(\nabla \Pi)^2}{\Pi} +\Pi(\nabla \Omega)^2\right]  + \\
    &  +\frac{\alpha_1 + \alpha_2}{4} \Pi^2 + \frac{\alpha_1 - \alpha_2}{4} (\Pi \cos \Omega)^2 -  \frac{1}{2}  \Pi \Delta \cos \Omega  -\delta \Pi \sin \Omega \cos \theta. \notag{}
\end{align}
\end{widetext}

Next, we compute the functional derivative of the Hamiltonian with respect to the field $\Omega$ to obtain its corresponding equation of motion:
\begin{widetext}
\begin{align}
   &\frac{\delta \mathscr{H}}{\delta \Omega} =
- \frac{1}{2} \Pi \sin\Omega\, \nabla \Theta  \nabla \theta
- \frac{1}{2} (\alpha_1 - \alpha_2) \Pi^2 \cos\Omega \sin\Omega
+\frac{1}{2} \Pi \Delta \sin\Omega - \delta \Pi \cos \Omega \cos \theta
- \frac{1}{4}\nabla (\Pi \nabla \Omega).
\end{align}
\end{widetext} 

Finally, taking into account the relaxation terms and substituting the expression for $\mathcal{Z}=\frac{1}{2}\Pi \cos \Omega$ everywhere, the system of equations of motion will take the form
\begin{widetext}
\begin{align}
\begin{cases}    \label{FinalSystem}
&\partial_t \Theta= \frac{1}{2} \left[ (\nabla \Theta)^2 +  \left( \frac{\nabla \theta}{2} \right)^2 \right] - \frac{1}{8} \left[ (\frac{\nabla \Pi}{\Pi}+ \nabla \Omega \cot{\Omega})^2 +\frac{(\nabla \Omega)^2}{\sin^2 \Omega} \right] + \\
&\qquad\qquad\qquad\qquad + \frac{\alpha_1+\alpha_2}{2}{\Pi} - \frac{1}{4} \nabla  \left[ \frac{\nabla \Pi}{\Pi}+\nabla \Omega \cot{\Omega} \right]  -\frac{\delta \cos \theta }{\sin \Omega} +\gamma_\Theta[\nabla(\Pi\nabla \Theta)+\frac{1}{2}\nabla(\Pi \cos \Omega \nabla \theta)]  ,\\ \\
&\partial_t \theta= \nabla \Theta \nabla \theta +(\alpha_1-\alpha_2)\Pi \cos \Omega-\Delta + \frac{1}{2} \frac{\nabla \Omega}{\sin \Omega} \left[ \frac{\nabla \Pi }{\Pi}+\nabla \Omega \cot{\Omega}  \right]  + \\
&\qquad\qquad\qquad\qquad +\frac{1}{2} \nabla  (\frac{\nabla \Omega}{\sin \Omega}) + 2 \delta \cot \Omega \cos \theta+ \gamma_\theta [\frac{1}{4}\nabla(\Pi\nabla \theta)+\frac{1}{2}\nabla(\Pi \cos \Omega \nabla \Theta)-\delta \Pi \sin \Omega \sin \theta] , \\
\\
&\partial_t \Pi=\nabla(\Pi\nabla \Theta)+\frac{1}{2}\nabla(\Pi \cos \Omega \nabla \theta),\\
\\
&\partial_t \Omega = \nabla \Omega  \nabla \Theta - \frac{\nabla  \Pi \nabla \theta}{2 \Pi} \sin \Omega -\frac{\nabla^2 \theta}{2} \sin \Omega-\frac{1}{2} \cos \Omega \nabla \Omega \nabla \theta +2 \delta \sin \theta + \\
&\quad  \quad \quad \quad \quad \quad \quad \quad  +\gamma_\Omega[\ \frac{1}{2} \Pi \sin\Omega\, \nabla \Theta  \nabla \theta
+ \frac{1}{2} (\alpha_1 - \alpha_2) \Pi^2 \cos\Omega \sin\Omega
- \frac{1}{2} \Pi \Delta \sin\Omega + \delta \Pi \cos \Omega \cos \theta
+ \frac{1}{4}\nabla (\Pi \nabla \Omega) ].
\end{cases}
\end{align}
\end{widetext}
The system \ref{FinalSystem} constitutes the central result, forming the foundation for all further analysis presented in this work.

\section{Spin relaxation in spatially homogeneous condensate and Landau-Lifshitz equation}

In the case of a spatially homogeneous condensate with zero momentum, the system of equations \ref{FinalSystem} simplifies significantly and takes the form:
\begin{widetext}
\begin{align}
\begin{cases}    \label{spathomogEq}
&\partial_t \Theta = \frac{\alpha_1+\alpha_2}{2} \Pi - \frac{\delta \cos \theta}{\sin \Omega},\\[8pt]
&\partial_t \theta = (\alpha_1 - \alpha_2)\Pi \cos \Omega - \Delta + 2\delta \cot \Omega \cos \theta - \gamma_\theta \delta \Pi \sin \Omega \sin \theta,\\[8pt]
&\partial_t \Pi = 0,\\[8pt]
&\partial_t \Omega = 2\delta \sin \theta + \gamma_\Omega \left[
\frac{1}{2}(\alpha_1 - \alpha_2) \Pi^2 \cos\Omega \sin\Omega
- \frac{1}{2} \Pi \Delta \sin\Omega + \delta \Pi \cos \Omega \cos \theta
\right].
\end{cases}
\end{align}
\end{widetext}

To represent the system in a more compact and physically transparent form, we introduce the Stokes vector $\mathbf{S} = (S_x, S_y, S_z)$, defined as:
\begin{align} 
\begin{cases}
S_x &= \Pi \sin \Omega \cos \theta, \\
S_y &= \Pi \sin \Omega \sin \theta, \\
S_z &= \Pi \cos \Omega.
\end{cases}
\end{align}

This vector characterizes the effective spin state of the two-component condensate and can be visualized on a generalized Bloch sphere. Taking the time derivatives of the Stokes vector components and using the second and fourth equations of system~\ref{spathomogEq} (see Supplementary Materials for details), the spin dynamics of the condensate can be compactly expressed in the form of a generalized Landau–Lifshitz–Gilbert (LLG) equation:
\begin{align} \label{LLG}
\partial_t \mathbf{S} = 
\mathbf{B}_\text{eff}(\mathbf{S}) \times \mathbf{S} 
- \lambda\, \mathbf{S} \times \left[\mathbf{B}_\text{eff}(\mathbf{S}) \times \mathbf{S}\right] 
+ \mathbf{F}_\text{aniso}(\mathbf{S}).
\end{align}
Here,
\begin{itemize}[label={}, leftmargin=*, nosep]
  \item $\lambda = \gamma_\Omega / 2$ is the effective damping coefficient;
  \item $\mathbf{B}_\text{eff}(\mathbf{S})$ is the \textit{effective magnetic field}, which is spin-dependent due to nonlinear interactions:
  \[
  \mathbf{B}_\text{eff}(\mathbf{S}) = 
  \begin{pmatrix}
  -2\delta \\
  0 \\
  -\Delta + (\alpha_1 - \alpha_2) S_z
  \end{pmatrix};
  \]
  \item $\mathbf{F}_\text{aniso}(\mathbf{S})$ is a force originating from \textit{anisotropic dissipation}, given by:
  \[
  \mathbf{F}_\text{aniso}(\mathbf{S}) = \delta \left( \gamma_\theta - \gamma_\Omega \frac{S^2}{S_\perp^2} \right)
  \begin{pmatrix}
  S_y^2 \\
  - S_x S_y \\
  0
  \end{pmatrix},
  \]
  where $S_\perp^2 = S_x^2 + S_y^2$ is the squared transverse spin component.
\end{itemize}

The first term in \ref{LLG} describes the coherent \textit{precession} of the spin vector $\mathbf{S}$ in the effective field $\mathbf{B}_\text{eff}$, analogous to the Larmor precession. The second term, proportional to $\lambda$, accounts for \textit{Gilbert damping}, driving the spin vector toward alignment with the effective field. The third term represents \textit{anisotropic dissipation}, which introduces angular-dependent relaxation and breaks axial symmetry. Thus, equation \ref{LLG} captures the dynamics of a nonlinear spin-like degree of freedom subject to both conservative precession and dissipative effects. The presence of spin-dependent interactions and anisotropic losses makes this a generalized nonlinear extension of the classical Landau–Lifshitz–Gilbert equation.

\section{Dynamics of the spatially uniform states}

Throughout the paper, we use a physically motivated dimensionless normalization. The conversion to physical units is as follows. Time is scaled by $t_{phys} = t\cdot \hbar / (g \rho)$, where $g = 6 \times 10^{-3} \, \text{meV} \cdot \mu\text{m}^2$, $\rho = 50 \, \mu\text{m}^{-2}$, and $\hbar = 6.582 \times 10^{-13} \, \text{meV} \cdot \text{s}$. The spatial coordinate is scaled by 
$x_{phys} = x \cdot \sqrt{ \hbar_C t_{phys} / m^* }$, where $\hbar_C = \hbar \cdot q_e$, $q_e = 1.6 \times 10^{-19} \, \text{C}$, and $m^* = 5 \times 10^{-5} m_e$, with $m_e = 9.109 \times 10^{-31} \, \text{kg}$. 
The wavevector is then $k_{\text{phys}} = k / x_0$, and the dimensional relaxation coefficient is given by $\gamma_{phys} = (2 \gamma / \rho) \cdot (x_{phys}^2 / t_{phys})$. The dimensionless parameters $\Delta$ and $\delta$ are converted to energy units by multiplying by $g \rho$: $\Delta_{phys} = \Delta \cdot g \rho$, $\delta_{phys} = \delta \cdot g \rho$.
This normalization allows direct comparison with experimentally relevant values.

The simplest yet physically significant case corresponds to the dynamics of the spatially uniform states. In this section, we use equations~\ref{LLG} to describe the temporal evolution of such states. We start with the case where the anisotropy term, characterized by $\delta$, dominates over Zeeman splitting, nonlinear effects, and relaxation terms. 
In the leading-order approximation, one readily identifies symmetric stationary solutions of the form
$\frac{\vec S_0}{|S_0|}=(\pm 1, 0, 0)^T$, where $|S_0|^2$ denotes the squared norm of the Stokes vector.

Analytical results can be easily obtained for small magnetic field $\Delta$, relaxation constants $\gamma_{\theta}$, $\gamma_{\Omega}$, and nonlinearity $\alpha_1$, $\alpha_2$. Then we seek a solution in the form $\vec S= (S_0, 0, 0)^T + (0, \xi_2, \xi_3)^T$, where $\xi_{2,3}$ are corrections to the second and third components of the Stokes vector. The equations for $\vec \xi=(\xi_2, \xi_3)^T$ read
\begin{align} \label{perturb_0}
\partial_t \vec \xi=\hat L \vec \xi -\vec f,  
\end{align}
where $\hat L= \begin{pmatrix}
  -\gamma_{\theta}S_0 \delta  &&  2\delta +(\alpha_1 - \alpha_2) S_0 \\
  -2 \delta  && -\frac{1}{2}\gamma_{\Omega}S_0 \left(2\delta + (\alpha_1-\alpha_2)S_0\right)
  \end{pmatrix}$ and 
   $\vec f= (\Delta S_0, -\frac{1}{2}\gamma_{\Omega}S_0^2 \Delta)^T$.

\begin{figure}[h!]
\begin{center}
\includegraphics[width=1.0\linewidth]{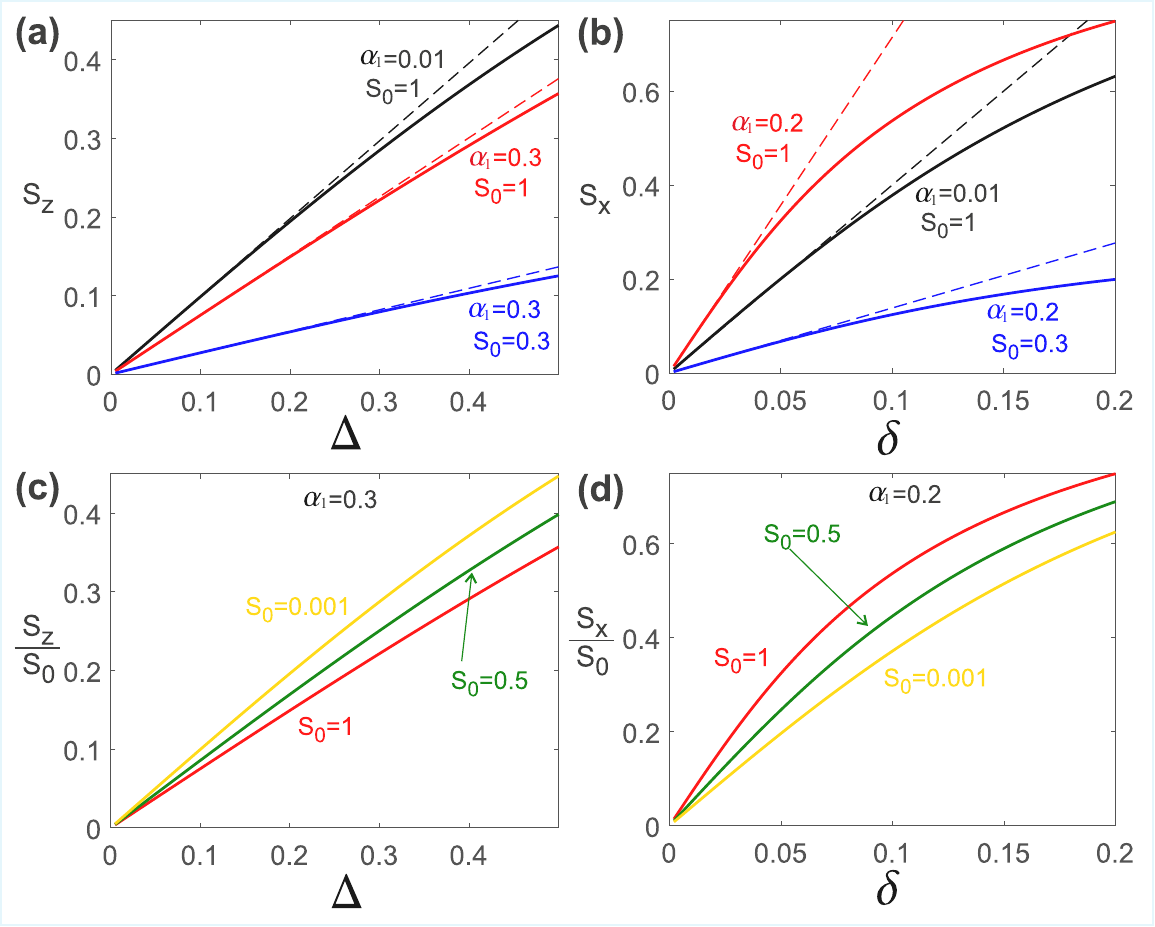}
\end{center}
\caption{The dependencies of the Stokes parameters $S_z$ on the magnetic field $\Delta$ are shown in panel (a) for different values of the repelling polariton-polariton interaction $\alpha_1=0.01$ (black) and $\alpha_1=0.3$ (red), the condensate density corresponds to $S_0=1$. The blue curve shows the dependency of $S_z$ for $\alpha_1=0.3$ and $S_0=0.3$. The anisotropy parameter for these curves is $\delta=0.5$. The solid curves are the numerically found solutions, while the dashed curves show the approximate solution \ref{Sz_1}.  Panel (b) shows the dependencies of $S_x$ on the anisotropy parameter $\delta$ for different values of the nonlinearity  $\alpha_1=0.01$ and $\alpha_1=0.2$, the Stokes vector length is $S_0=1$. The blue curve corresponds to $\alpha_1=0.2$, $S_0=0.3$  The curves are plotted for the magnetic field $\Delta=0.5$. The solid curves represent numerically found solutions; the dashed curves are the approximate solutions given by \ref{xi_nl1}. Panels (c) and (d) show the dependencies of the normalized $z$ and $x$ components of the Stokes vector  $\frac{S_z}{S_0}$, $\frac{S_x}{ S_0}$ on the magnetic field $\Delta$ and anisotropy $\delta$ for different densities of the condensate. The polariton-polariton repelling interaction is $\alpha_1=0.3$ for panel (c) and $\alpha_1=0.2$ for panel (d). The other parameters are  $\alpha_2=-\alpha_1$, $\gamma_{\Omega}=0.5$ and $\gamma_{\theta}=0.2$ for all panels. \label{fig0}}
\end{figure}

A straightforward solution to~\ref{perturb_0} is given by $\vec \xi= \left( 0, \frac{S_0 \Delta }{2\delta +(\alpha_1-\alpha_2)S_0} \right)^T$. So we can conclude that the introduction of the magnetic field results in the appearance of the non-zero component 
\begin{align} \label{Sz_1}
S_z=\frac{S_0 \Delta }{2\delta +(\alpha_1-\alpha_2)S_0}
\end{align}
of the Stokes vector,  while $S_y$ remains zero. In the following, we show that the state is stable if $\delta (\alpha_1-\alpha_2)S_0>0$. From \ref{Sz_1} it is seen that for the stable state the nonlinearity reduces the absolute value of $S_z$, see Fig.~\ref{fig0}. 

To study the linear stability of the equilibrium points, we need to find the eigenvalues of $\hat L$. This can be done in the general case, but for the sake of transparency, we assume that the effect of the relaxation is small compared to that of magnetic fields and nonlinearity.  We can represent $\hat L$ as $\hat L=\hat L_c + \hat L_d$, where \mbox{$\hat L_c= \begin{pmatrix}
  0  &&  2\delta +(\alpha_1 - \alpha_2) S_0 \\
  -2 \delta  && 0
  \end{pmatrix}$}
accounts for the conservative dynamics and
\mbox{$\hat L_d= \begin{pmatrix}
  -\gamma_{\theta}S_0 \delta  &&  0 \\
  0  && -\frac{1}{2}\gamma_{\Omega}S_0 \left(2\delta + (\alpha_1-\alpha_2)S_0\right)
  \end{pmatrix}$} describes the relaxation. Treating $\hat L_r$ as a small perturbation, we can find the eigenfrequencies $\omega$
  \begin{widetext}
\begin{align} \label{perturb_stab}
\omega = \pm \sqrt{2\delta(2\delta +(\alpha_1-\alpha_2)S_0 )} - \frac{i}{2} \left( \gamma_{\theta} +\gamma_{\Omega} \left( 1+\frac{\alpha_1-\alpha_2}{2}\frac{S_0}{\delta} \right) \right)\delta S_0.
\end{align}
\end{widetext}

Thus, the stability analysis reveals that solutions with $\delta S_0 > 0$ are always linearly stable, provided the condition $\alpha_1 - \alpha_2 > 0$ is satisfied. In contrast, solutions characterized by $\delta S_0 < 0$ exhibit linear instability under the same condition. It should be emphasized that the regime $\alpha_1 - \alpha_2 > 0$ is typical for exciton-polariton condensates, where $\alpha_1 > 1$ and $\alpha_1 \alpha_2 < 0$. Furthermore, in the case of attractive interactions ($\alpha_1 < 1$), spatially homogeneous solutions are modulationally unstable and, therefore, of limited physical relevance. It should be noted that there are symmetry-broken solutions, but for the repelling interaction $\alpha_1>0$ they are unstable.

As expected, relaxation is a nonlinear process and, consequently, the decay time to a stationary state depends on the length of the Stokes vector $S_0$ (and thus on the density of the condensate) even in the absence of the conservative polariton-polariton interaction $\alpha_1=\alpha_2=0$. The conservative polariton-polartion interaction affects the relaxation rate $\Im \omega$ but for the repelling interaction $\alpha_1>0$ it does not change the sign of $\Im \omega$ and hence does not affect the stability of the states. 

Now, let us examine the case of a relatively strong magnetic field and consider anizotropy, nonlinearity, and relaxation as perturbations. Then in the leading approximation order, the solution is $\vec S = (0, 0, S_0)^T$. We search for a solution of the form $\vec S =(0, 0, S_0)^T+(\xi_1, \xi_2, 0)^T$. 
 \begin{widetext}
\begin{align} \label{perturb_xi_nl1}
\dot \xi_1=(\Delta-(\alpha_1-\alpha_2)S_0 ) \xi_2 - \frac{\gamma_{\Omega}}{2}\left( S_0 (\Delta-(\alpha_1-\alpha_2)S_0 ) \xi_1 -2\delta S_0^2\right) - \delta \gamma_{\Omega}S_0^2 \frac{\xi_2^2}{\xi_1^2+\xi_2^2} \\
\dot \xi_2=((\alpha_1-\alpha_2)S_0-\Delta ) \xi_1 - \frac{\gamma_{\Omega}}{2} S_0 (\Delta-(\alpha_1-\alpha_2)S_0 )\xi_2+2\delta S_0 + \delta \gamma_{\Omega}S_0^2 \frac{\xi_1 \xi_2}{\xi_1^2+\xi_2^2}. \label{perturb_xi_nl2}
\end{align}
\end{widetext}
Although the nonlinear terms are perturbatively small, a linearization around the origin $(0,0)$ is not valid, since the terms involved are nondifferentiable at that point. Nevertheless, the solutions can be found explicitly:
\begin{align}
\xi_1 &= \frac{2\delta S_0}{\Delta - (\alpha_1 - \alpha_2)S_0} \label{xi_nl1} \\
\xi_2 &= 0. \label{xi_nl2}
\end{align}
So, the introduction of anisotropy results in the appearance of finite $S_x=\frac{2\delta S_0}{\Delta-(\alpha_1-\alpha_2)S_0}$, but $S_y$ remains zero. Let us remark that in developing the perturbation theory we assumed $|\Delta| \gg (\alpha_1-\alpha_2) S_0$ and therefore the denominator in \ref{xi_nl1} is never zero within the validity limit of the theory.

To examine the stability of the equilibrium points, we can linearize the equations \ref{perturb_xi_nl1}, \ref{perturb_xi_nl2} at the equilibrium point \ref{xi_nl1}, \ref{xi_nl2}. The deviation $\vec \eta =(\eta_1, \eta_2)^T$ from the equilibrium point is  described by the equation 
\begin{align} \label{perturb_1}
\partial_t \vec \eta=\hat L \vec \eta,  
\end{align}
where $$\hat L= \begin{pmatrix}
  -\frac{1}{2}\gamma_{\Omega} S_0 (\Delta-(\alpha_1-\alpha_2)S_0)  &&  \Delta-(\alpha_1-\alpha_2)S_0 \\
  (\alpha_1-\alpha_2)S_0-\Delta  && 0
  \end{pmatrix}.$$ The eigenvalues of $\hat L$ define the stability of the equilibrium point. The eigenfrequencies are given by  
  \begin{align} \label{stab2}
\omega = \frac{\Delta-(\alpha_1-\alpha_2)S_0 }{4} \left(  \pm \sqrt{ 16-\gamma_{\Omega}^2 S_0^2   }  - i \gamma_{\Omega} S_0   \right).
\end{align}
From \ref{stab2} we can conclude that the relaxation rate to this equilibrium point depends on the density of the condensate. The polariton-polariton interaction affects the relaxation rate, but this interaction cannot change the sign of the imaginary part of the eigenfrequency $\Im \omega$ for the repelling interaction $\alpha_1>0$ and so the interaction does not affect the stability of the equilibrium point.

\begin{figure}[t!]
\includegraphics[width=1.0\linewidth]{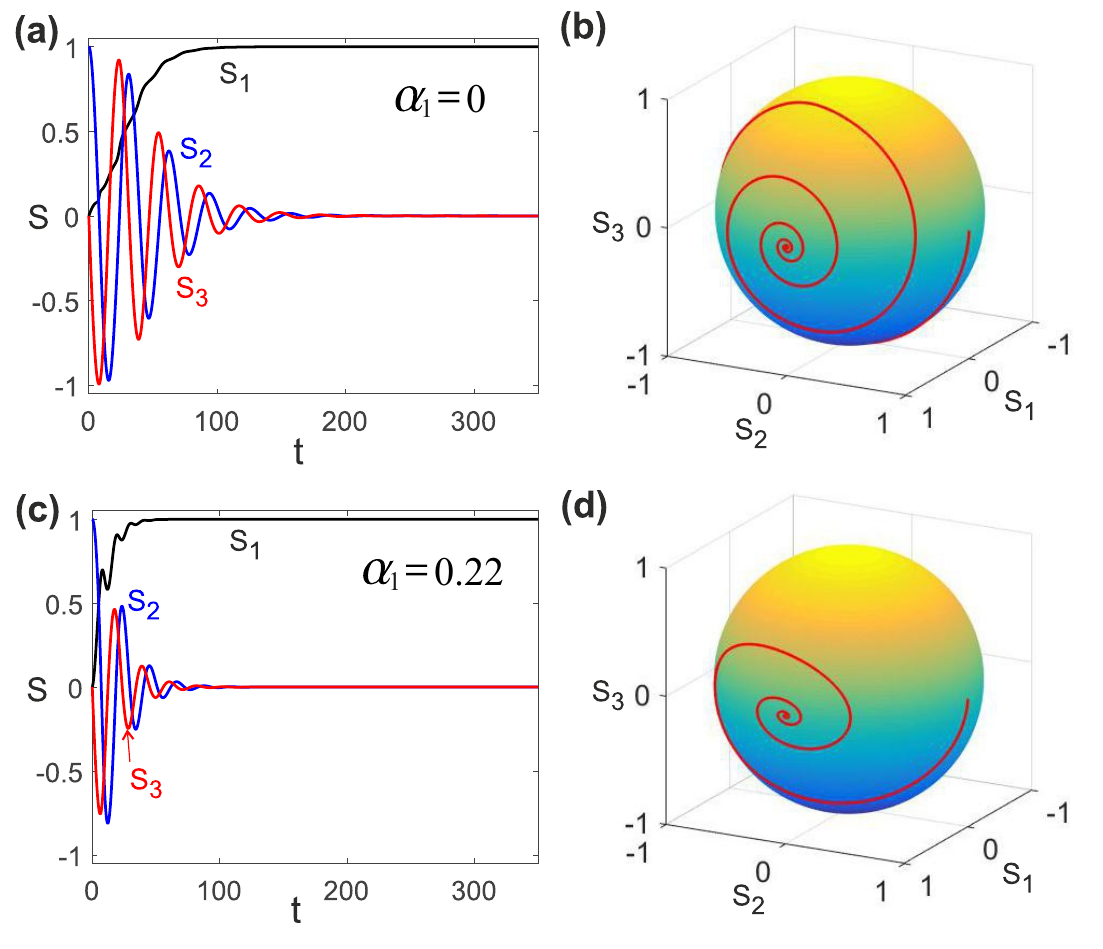}
\caption{ The temporal evolution of the Stokes parameters are shown in panels (a) and (c) for $\alpha_1=0$ and $\alpha_1=0.22$ correspondingly. Panels (b) and (d) show the motion of the Stokes vector on the Poincare sphere. The parameters are $ \alpha_2 = -0.1 \alpha_1$, $\Delta=0$, $\delta=0.1$, $\gamma_{\Omega}=0.5$, $\gamma_{\theta}=0.2$.\label{fig1} }
\end{figure}

The dynamics of the Stokes parameters illustrating the relaxation to the stable states are shown in Fig.~\ref{fig1} for a zero magnetic field $\Delta=0$. Comparing the upper and lower rows plotted for $\alpha_1=0$ and $\alpha_1=0.22$ we can conclude that, according to \ref{perturb_stab}, the relaxation to the equilibrium point is faster in the presence of nonlinearity. It is also seen that the nonlinearity affects the shape of the Stokes vector trajectory.

\begin{figure}[t!]
\includegraphics[width=1.0\linewidth]{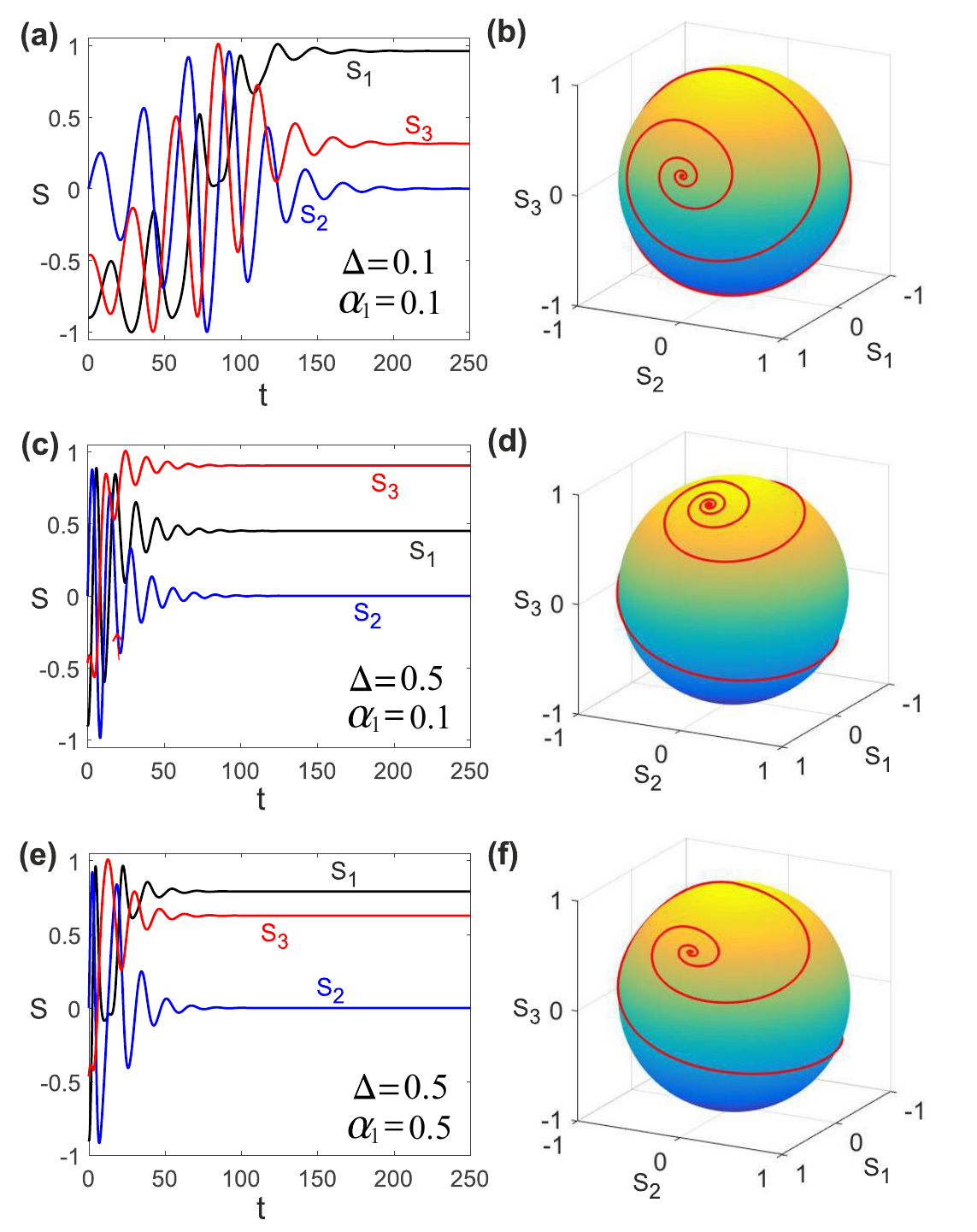}
\caption{ The temporal evolution of the Stokes parameters are shown in panels (a) and (c) for $\alpha_1=0.1$ and $\Delta=0.1$, $\Delta=0.5$ correspondingly. Panel (e) shows the temporal evolution of the Stokes parameters for $\alpha_1=0.5$ and $\Delta=0.1$, $\Delta=0.5$. The motion of the Stokes parameter on the Poincare sphere for all these cases is shown in panels (b), (d) and (f). The other parameters are  $\delta=0.1$, $\gamma_{\Omega}=0.5$, $\gamma_{\theta}=0.2$.\label{fig2} }
\end{figure}

Relaxation toward the equilibrium point in the presence of an external magnetic field is illustrated in Fig.~\ref{fig2}. As seen from the figure, the \( S_y \) component of the Stokes vector remains identically zero throughout the evolution, in full agreement with the analytical results. For relatively weak magnetic fields [panels (a) and (b)], the \( S_z \) component remains small, indicating that the equilibrium lies close to the equatorial plane. As the magnetic field increases, the \( S_z \) (or \( S_3 \)) component grows, and the equilibrium point shifts toward the pole---see the transition from panels (a), (b) to (c), (d).

An increase in the nonlinearity parameter \( \alpha_1 \) also affects the equilibrium configuration: it moves the equilibrium point away from the pole, as shown in panels (e) and (f). A comparison of the relaxation rates further reveals that relaxation proceeds faster in stronger magnetic fields, as evidenced by the contrast between Fig.~\ref{fig2}(a) and Fig.~\ref{fig2}(c). This agrees with our analytical predictions. 

\section{Dispersion of elementary excitations}

To gain insight into the spectrum of collective modes, we derive and analyze the dispersion of elementary excitations. A detailed calculation of all intermediate steps can be found in the Supplementary Materials. As a starting point, we consider the idealized case without relaxation, i.e., we set $\gamma_\Omega = \gamma_\theta = \gamma_\Theta = 0$ in the system~\ref{FinalSystem}.

We look for stationary, spatially homogeneous configurations that minimize the energy functional. Such equilibrium states, denoted by $(\Pi_0,\, \Theta_0,\, \theta_0,\, \Omega_0)$, are defined by the following conditions:
\begin{equation}
\begin{cases} \label{stationarystate}
    \partial_t \Pi = 0,\\
    \partial_t \Omega = 0,\\
    \partial_t \theta = 0,\\
    \partial_t \Theta = \mu,
\end{cases}
\quad \text{evaluated at } (\Pi_0, \Theta_0, \theta_0, \Omega_0),
\end{equation}
where $\mu$ is the chemical potential. From~\ref{stationarystate}, one obtains the constraint $\sin \theta_0 = 0$, which implies that the equilibrium value of the relative phase is $\theta_0 = 0$, that is, the two condensate components are phase-locked. Substituting this into the remaining equations in~\ref{stationarystate} allows us to derive the condition for the equilibrium polar angle $\Omega_0$:
\begin{equation} \label{Omega_0}
(\alpha_1 - \alpha_2)\Pi_0 \cos \Omega_0 - \Delta + 2\delta \cot \Omega_0 = 0.
\end{equation}

Furthermore, we obtain the expression for the chemical potential $\mu$ corresponding to this stationary state:
\begin{equation} \label{mu}
\mu = \frac{\alpha_1 + \alpha_2}{2} \Pi_0 - \frac{\delta}{\sin \Omega_0}.
\end{equation}

To derive the spectrum of elementary excitations, we linearize the system around the stationary state. Introducing small fluctuations,
\begin{equation} \label{ansatz}
\begin{cases} 
    \Pi = \Pi_0 + \delta\Pi(\bm{r},t), \\
    \Theta = \mu t + \delta\Theta(\bm{r},t), \\
    \theta = \theta_0 + \delta\theta(\bm{r},t), \\
    \Omega = \Omega_0 + \delta\Omega(\bm{r},t),
\end{cases}
\end{equation}
with $|\delta f| \ll |f_0|$, we substitute~\ref{ansatz} into the non-dissipative equations~\ref{FinalSystem} and linearize in $\delta \Pi$, $\delta \Theta$, $\delta \theta$, and $\delta \Omega$.

To extract the dispersion relations, we assume plane-wave perturbations of the form
\begin{equation} \label{perturb}
\delta f(\mathbf{r}, t) = A_f \cos(\mathbf{k} \cdot \mathbf{r} - \omega t) + B_f \sin(\mathbf{k} \cdot \mathbf{r} - \omega t),
\end{equation} 
where $f \in \{ \Pi, \Theta, \theta, \Omega \}$. Substituting these into linearized equations yields a homogeneous linear system for the coefficients $\{A_f, B_f\}$, which can be written in matrix form as $\mathbb{M}(\omega, \mathbf{k}) \cdot \vec{v} = 0$. The condition for nontrivial solutions,
\[
\det \mathbb{M}(\omega, \mathbf{k}) = 0,
\]
defines the dispersion relation $\omega(\mathbf{k})$ of the elementary excitations.

In the case $\delta = 0$ and $\Delta \neq 0$, which corresponds to a symmetric cavity placed in the external magnetic field along the z axis, the dispersion relation takes the form:
\begin{align}
    \omega^2 = \omega_0^2 + \Pi_0 \left[ \alpha_1 \pm \sqrt{ \alpha_2^2 + \frac{ \alpha_1^2 - \alpha_2^2 }{4\omega_c^2} \Delta^2 } \right] \omega_0,
\end{align}
where $\omega_0 = \frac{k^2}{2}$ is the bare polariton dispersion with $\omega_0(0) = 0$, and we define the characteristic interaction frequency $\omega_c = \frac{\alpha_1 - \alpha_2}{2} \Pi_0$.

In contrast, for $\delta \neq 0$ and $\Delta = 0$, which correspond to a system with polarization splitting but no external magnetic field, the excitation spectrum splits into two branches:
\begin{align}
\begin{cases}
    \omega_+^2 = \omega_0^2 + \omega_0 \Pi_0 (\alpha_1 + \alpha_2), \\
    \omega_-^2 = (\omega_0 + 2\delta)^2 + 2\omega_c (\omega_0 + 2\delta).
\end{cases}
\end{align}
In this regime, the $\omega_+$ branch corresponds to the density (Bogoliubov-like) mode, while the $\omega_-$ branch is associated with polarization (spin-wave-like) dynamics. The coupling between polarization degrees of freedom and interactions leads to a finite gap in the spectrum even at zero momentum:
\begin{align}
    E_g = \sqrt{4\delta \omega_c + 4\delta^2}.
\end{align}
This gap reflects the energy cost of exciting a relative polarization oscillation in the condensate, and is tunable via both the interaction strength and the polarization splitting parameter $\delta$.

\begin{figure}[h!]
\includegraphics[width=1.0\linewidth]{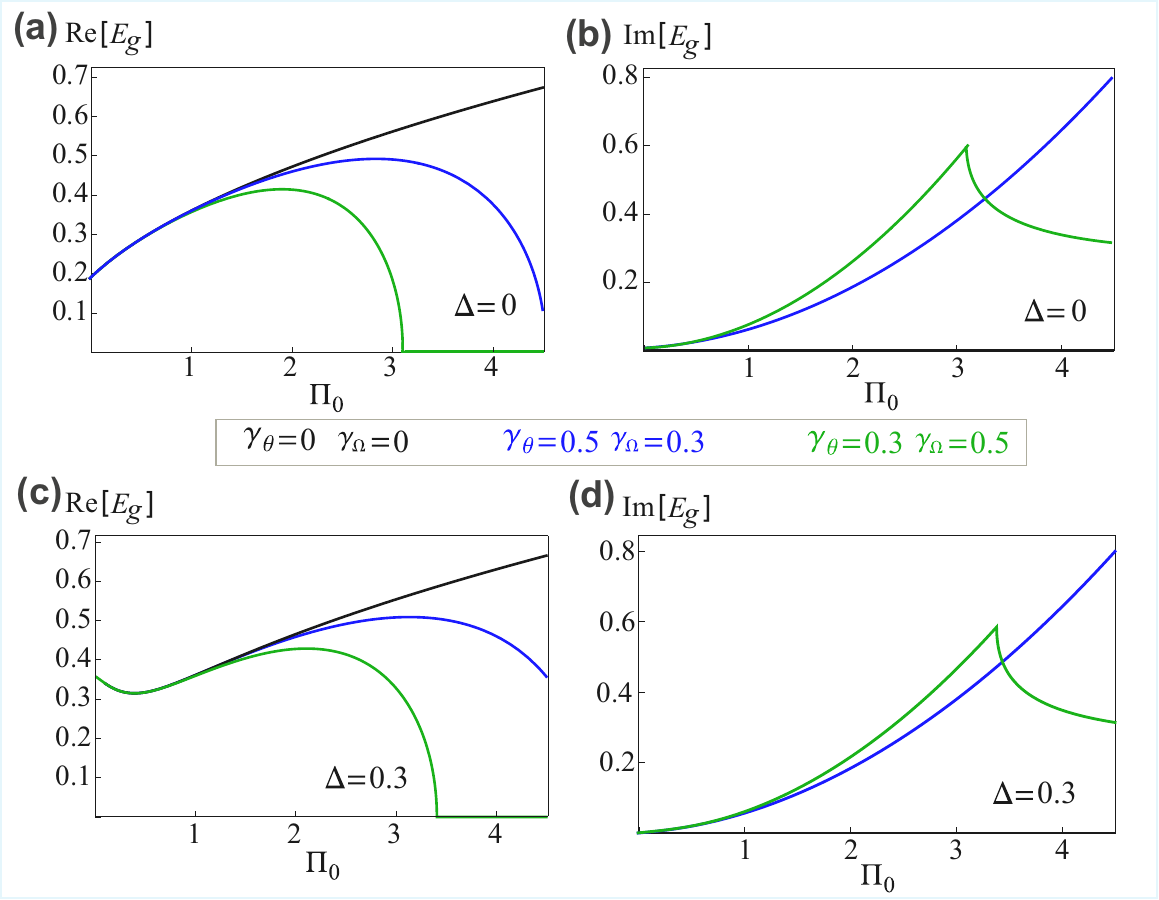}
\caption{Dependence of the real and imaginary parts of the excitation gap on the condensate density \(\Pi_0\) for different values of the relaxation parameters. The black curves correspond to the relaxation-free case, while the blue and green ones represent the scenarios with \(\gamma_\theta = 0.5,\ \gamma_\Omega = 0.3\) and \(\gamma_\theta = 0.3,\ \gamma_\Omega = 0.5\), respectively. Panels (a) and (b) illustrate the situation without Zeeman splitting (\(\Delta = 0\)), whereas panels (c) and (d) correspond to the case with an external magnetic field inducing Zeeman splitting of \(\Delta = 0.3\). The remaining parameters are fixed as \(\alpha_1 = 0.5\), \(\alpha_2 = -0.1\alpha_1\), and \(\delta = 0.1\). \label{gap3}}
\end{figure}

These results are in exact agreement with those obtained in~\cite{Shelykh2007}. The general case with both $\delta \neq 0$ and $\Delta \neq 0$, while neglecting relaxation, yields more complex analytical expressions. The full dispersion relations for this regime are presented in the Supplementary Materials. For completeness, we only present the expression for the energy gap in the general case:

\begin{align} \label{E_g}
E_g=  \sqrt{4\delta \omega_c \sin \Omega_0+\frac{4\delta^2}{\sin^2\Omega_0}},
\end{align}
where $\Omega_0$ satisfies the stationary condition given in equation~\ref{Omega_0}.

Next, we calculate the dispersion of elementary excitations taking into account the relaxation terms. We consider the system~\ref{FinalSystem} and follow the same approach as previously discussed. From system~\ref{FinalSystem}, we recover the same stationary state conditions as those given by $\theta_0 = 0$ and the equations~\ref{Omega_0} and~\ref{mu}. This is expected since relaxation does not alter the fundamental energy minimum of the system but only influences the dynamical evolution as the system approaches this minimum. The next step involves linearizing the system around its equilibrium state by introducing plane-wave perturbations as in~\ref{perturb}, which yields a matrix $\mathbb{M}$ whose determinant defines the dispersion relation. Due to the considerable complexity of the resulting expression, we analyze the obtained dispersion numerically. 
To investigate the formation and evolution of the excitation gap, we begin by analyzing the behavior of the dispersion relation at a zero-wave vector. In this limit, the characteristic equation, obtained from the condition of the vanishing determinant of the linearized system, yields five solutions: a four-fold degenerate zero-frequency solution and a pair of non-zero symmetric roots \(\pm \omega_+\) and \(\pm \omega_-\). The latter are given by:
\begin{widetext}
\begin{equation}
\omega_\pm^2 =
4\delta \omega_c \sin \Omega_0 + \frac{4\delta^2}{\sin^2 \Omega_0}
- 2a^2 - 2b^2 \pm 2(-a + b) \sqrt{(a + b)^2 - \left(4\delta \omega_c \sin \Omega_0 + \frac{4\delta^2}{\sin^2 \Omega_0}\right)},
\end{equation}
\end{widetext}
where the parameters \(a\) and \(b\) take the form:
\begin{equation*}
a = \frac{\gamma_\theta}{2} \Pi_0 \delta \sin \Omega_0, \qquad
b = \frac{\gamma_\Omega}{2} \Pi_0 \left( \omega_c \cos 2\Omega_0 - \frac{\Delta}{2} \cos \Omega_0 \right).
\end{equation*}

In what follows, we define the excitation gap as the lowest of the two non-zero frequencies, namely \(\omega_-\). This quantity represents the minimal energy required to excite a mode above the condensate background.

\begin{figure}[h!]
\includegraphics[width=1.0\linewidth]{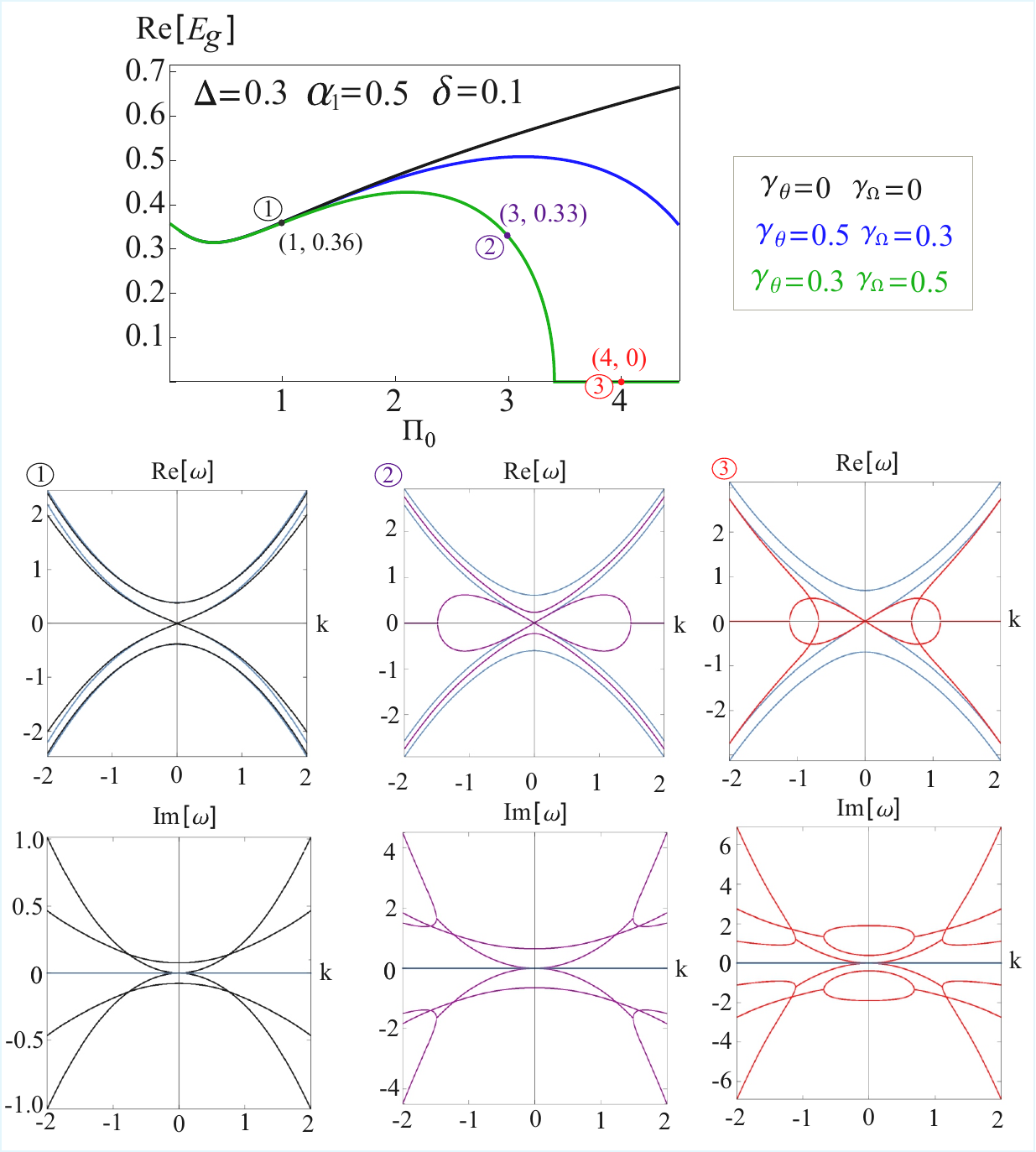}
\caption{The upper panel presents the dependence of the real part of the excitation gap on the condensate density \(\Pi_0\) for several relaxation scenarios. The black curve corresponds to the relaxation-free case, while the blue and green curves represent cases with relaxation parameters \(\gamma_\theta = 0.5,\ \gamma_\Omega = 0.3\) and \(\gamma_\theta = 0.3,\ \gamma_\Omega = 0.5\), respectively. Three representative points on the green curve are selected to illustrate distinct dispersion regimes: the first point \((1,\, 0.36)\) is marked in black, the second \((3,\, 0.33)\) in purple, and the third \((4,\, 0)\) in red. For each of these density values, the corresponding real and imaginary parts of the excitation spectrum are shown in the lower panels. In each dispersion plot, the bold curve describes the relaxation-influenced spectrum at the selected density (indicated by the color matching the marked point on the upper panel), while the light blue curve shows the corresponding spectrum in the absence of relaxation. This visual comparison emphasizes the qualitative changes induced by dissipative effects. The parameters used in all plots are fixed as \(\alpha_1 = 0.5\), \(\alpha_2 = -0.1\alpha_1\), \(\gamma_\Theta = 0.5\), and \(\delta = 0.1\).\label{dispersion}}
\end{figure} 

Next, we examine the real and imaginary parts of \(\omega_-\) as functions of key system parameters. In the main text, we restrict ourselves to the dependence on the polariton density \(\Pi_0\) for several representative values of the relaxation coefficients, while additional parameter dependencies are provided in the Supplementary Materials.

Figure~\ref{gap3} illustrates the evolution of the excitation gap as a function of \(\Pi_0\) in various dissipation regimes. The black curves correspond to the non-relaxation limit \((\gamma_\theta = \gamma_\Omega = 0)\), in which the gap remains purely real. Colored curves represent cases with finite relaxation. As expected, the introduction of dissipation leads to a nonzero imaginary part of the excitation frequency, signifying damping or potential instabilities. Moreover, increasing the density tends to reduce the critical point at which the gap closes. The application of an external magnetic field \(\Delta \neq 0\) (panels (c) and (d)) shifts this critical density to higher values, thereby stabilizing the system.

To further elucidate the effect of relaxation, we select three representative points from the density dependence and analyze the corresponding dispersion profiles, see figure~\ref{dispersion}. Comparison of the spectra before (pale blue lines) and after (bold colored lines) the inclusion of dissipation reveals that relaxation fundamentally alters the dispersion landscape. Notably, the excitation gap may collapse at high densities --- a phenomenon absent in the conservative regime.

An analysis of other parameter dependencies (see Supplementary Materials) reveals a particularly interesting behavior: increasing \(\gamma_\Omega\) at fixed \(\gamma_\theta = 0.3\) can also induce gap closure, while increasing \(\gamma_\theta\) at fixed \(\gamma_\Omega = 0.3\) does not exhibit this tendency. This asymmetry underscores the qualitatively different roles played by the two relaxation channels in the stability and excitation spectrum of the system.

\section{Conclusion}

In this work, we derived the dynamical equations governing the evolution of exciton-polariton condensates with a polarization (spin) degree of freedom, explicitly incorporating a spin relaxation term. While conserving the total number of particles, this term allows for spin interconversion processes that reduce the total energy of the system, thereby driving it toward a local energy minimum. This formulation enables a consistent treatment of dissipative spin dynamics, which are essential for the description of nonequilibrium condensate behavior.

The resulting equations take the form of hydrodynamic equations for two interacting quantum fluids and can be viewed as a generalized extension of the vector Gross-Pitaevskii equations commonly used in the mean-field approximation of polariton condensates. Although, in principle, the system may be reformulated in terms of a generalized Gross-Pitaevskii equation, such a transformation is technically involved and tends to obscure the physical interpretation. Therefore, we conducted our analysis within the hydrodynamic approach, which offers a more transparent view of the underlying physics and facilitates analytical treatment.

To elucidate the impact of the relaxation term, we restricted our consideration to a spatially uniform condensate and examined the system’s evolution toward energetically favorable stationary states. A perturbative approach was developed to construct stationary solutions when either the external magnetic field or the polarization anisotropy can be treated as a small perturbation. Within this framework, we derived analytical expressions for the eigenvalues that determine the linear stability of these states. The validity of the analytical results was confirmed through numerical simulations, which also provided insights into the nonlinear dynamics beyond the perturbative regime.

In the final part of the work, we analyzed the dynamics of small excitations on top of spatially uniform backgrounds. By linearizing the hydrodynamic equations, we derived the dispersion relation for elementary excitations (bogolons). The analysis showed that the inclusion of relaxation leads to a finite imaginary component in the excitation spectrum, reflecting the dissipative nature of the dynamics. Moreover, we demonstrated that relaxation can lead to a closure of the energy gap between dispersion branches --- a behavior absent in the conservative limit.

We anticipate that the inclusion of the relaxation term will significantly improve the accuracy of theoretical descriptions of exciton-polariton condensates. The use of a relatively simple mean-field approximation provides both analytical clarity and computational efficiency, making it possible to investigate complex spatio-temporal dynamics that would be challenging to access using more intricate quantum kinetic equations. The interplay between energy relaxation and other dissipative effects --- such as particle loss due to absorption — constitutes a promising direction for future research, which lies beyond the scope of the present study.

\bibliography{main}   

\end{document}


\title{Supplemental Materials: \\ Spin relaxation in a polariton fluid: quantum hydrodynamic approach}

\author{D. A. Saltykova}
\affiliation{ITMO University, St. Petersburg 197101, Russia}

\author{A. V. Yulin}
\affiliation{ITMO University, St. Petersburg 197101, Russia}

\author{I. A. Shelykh}

\affiliation{Science Institute, University of Iceland, Dunhagi 3, IS-107 Reykjavik, Iceland}

\maketitle
\widetext
\tableofcontents

\section*{Supplemental Note 1: Derivation of equations}

\noindent

In this section, we present the derivation of the dynamical equations describing the coherent evolution of a spinor polariton condensate in the presence of external magnetic fields and coherent coupling. Our starting point is the system of coupled Gross–Pitaevskii-type equations describing the two circular polarization components of the condensate wave function, \mbox{$\psi_1$ and $\psi_2$:}
\begin{equation*}
    \begin{cases}
    i \partial_t \psi_1= - \frac{1}{2} (\partial_x^2+\partial_y^2)\psi_1 + \alpha_1|\psi_1|^2\psi_1+\alpha_2|\psi_2|^2\psi_1-\frac{\Delta}{2}\psi_1-\delta \psi_2, \\
    
    i \partial_t \psi_2= - \frac{1}{2} (\partial_x^2+\partial_y^2)\psi_2 +\alpha_1|\psi_2|^2\psi_2+\alpha_2|\psi_1|^2\psi_2+\frac{\Delta}{2}\psi_2-\delta \psi_1 
    \end{cases}
\end{equation*}
These equations can be derived from a variational principle using the Hamiltonian formalism. In particular, they can be cast into the compact form:
\begin{equation*}
    \begin{cases}
    i \partial_t \psi_1= \frac{\delta \mathscr{H}}{\delta \psi^{*}_1}, \\
    
    i \partial_t \psi_2=  \frac{\delta \mathscr{H}}{\delta \psi^{*}_2},
    \end{cases}
\end{equation*}
where $\delta \mathscr{H}/\delta \psi^*_{1,2}$ denotes the functional derivative of the Hamiltonian with respect to the complex conjugate fields. It has the general form:
\begin{equation*}
     \frac{\delta \mathscr{H}}{\delta \psi^{*}_{1,2}} =\frac{\partial\mathscr{H}}{\partial \psi_{1,2}^{*}}-\partial_\mu\left(\frac{\partial\mathscr{H}}{\partial(\partial_\mu\psi_{1,2}^{*})}\right)=\frac{\partial\mathscr{H}}{\partial \psi_{1,2}^{*}}-\partial_t\left(\frac{\partial\mathscr{H}}{\partial(\partial_t\psi_{1,2}^{*})}\right)-\partial_x\left(\frac{\partial\mathscr{H}}{\partial(\partial_x\psi_{1,2}^{*})}\right)-\partial_y\left(\frac{\partial\mathscr{H}}{\partial(\partial_y\psi_{1,2}^{*})}\right); 
\end{equation*}
with the index $\mu$ running over the space-time coordinates: $t$, $x$, and $y$.

To obtain the explicit form of the Hamiltonian density $\mathscr{H}$, we start from the Lagrangian density $\mathscr{L}$, which in our case reads:
\begin{equation*}
    \mathscr{L}= \frac{i}{2}(\psi_1^{*}\frac{\partial \psi_1}{\partial t} - \psi_1 \frac{\partial \psi_1^{*}}{\partial t})+  \frac{i}{2}(\psi_2^{*}\frac{\partial \psi_2}{\partial t} - \psi_2 \frac{\partial \psi_2^{*}}{\partial t})-\frac{1}{2} (| \nabla \psi_1|^2+|\nabla \psi_2|^2)- \frac{\alpha_1}{2}(|\psi_1|^4+|\psi_2|^4)-\alpha_2|\psi_2\psi_1|^2 +\frac{\Delta}{2}(|\psi_1|^2-|\psi_2|^2)+\delta(\psi_1 \psi_2^\ast+\psi_2 \psi_1^\ast), 
\end{equation*}
where $| \nabla \psi_{1,2}|^2=|\partial_x \psi_{1,2}|^2+|\partial_y \psi_{1,2}|^2$.
The corresponding Hamiltonian density is obtained via the Legendre transformation:
\begin{equation*}
     \mathscr{H}=\sum_{I=1,2}\frac{\partial \mathscr{L}}{\partial (\partial_t \psi_I)} \partial_t \psi_I +\sum_{I=1,2}\frac{\partial \mathscr{L}}{\partial (\partial_t \psi^{*}_I)} \partial_t \psi^{*}_I -\mathscr{L}.
\end{equation*}
After straightforward calculation, one obtains:
\begin{equation*}
    \mathscr{H}= \frac{1}{2} (| \nabla \psi_1|^2+|\nabla \psi_2|^2)+ \frac{\alpha_1}{2}(|\psi_1|^4+|\psi_2|^4)+\alpha_2|\psi_1 \psi_2|^2 - \frac{\Delta}{2}(|\psi_1|^2-|\psi_2|^2)-\delta(\psi_1 \psi_2^\ast+\psi_2 \psi_1^\ast).
\end{equation*}

In order to proceed toward a hydrodynamic representation and eventually introduce dissipative terms, we express the complex wave functions in the Madelung form:
\begin{equation*}
\begin{cases}
    \psi_1(\vec{r},t) = \sqrt{\rho_1(\vec{r},t)} e^{-i \theta_1(\vec{r},t)}, \\
     \psi_2(\vec{r},t) = \sqrt{\rho_2(\vec{r},t)} e^{-i \theta_2(\vec{r},t)},
     \end{cases}
\end{equation*}
where $\rho_{1,2}(\vec{r},t)$ are real-valued densities and $\theta_{1,2}(\vec{r},t)$ are the corresponding real-valued phases of the spin components. Substituting these expressions into the Lagrangian density, we obtain the following real-field form of the Lagrangian:
\begin{align*}
\mathcal{L} =\ 
&  \rho_1 \partial_t \theta_1 +\rho_2 \partial_t \theta_2  - \frac{1}{2} \left[ \rho_1 (\nabla \theta_1)^2 + \rho_2 (\nabla \theta_2)^2 \right] 
- \frac{1}{2} \left[ \frac{(\nabla \rho_1)^2}{4\rho_1} + \frac{(\nabla \rho_2)^2}{4\rho_2} \right]   -\\ -&\frac{\alpha_1}{2} (\rho_1^2 + \rho_2^2)- \alpha_2 \rho_1 \rho_2 + \frac{\Delta}{2} (\rho_1 - \rho_2) +2 \delta \sqrt{\rho_1\rho_2} \cos(\theta_1-\theta_2).
\end{align*}

We identify the canonically conjugate momenta associated with the phase fields as follows:
\begin{equation*}
\begin{cases}
    \pi_1=\frac{\partial\mathscr{L}}{\partial(\partial_t\theta_1)}=\rho_1, \\
    \pi_2=\frac{\partial\mathscr{L}}{\partial(\partial_t\theta_2)}=\rho_2.
     \end{cases}
\end{equation*}

This allows us to construct the Hamiltonian density in terms of $\rho_{1,2}$ and $\theta_{1,2}$:
\begin{align*}
     &\mathscr{H}=\pi_1\partial_t\theta_1+\pi_2\partial_t\theta_2 -\mathscr{L}=  \frac{1}{2} \left[ \rho_1 (\nabla \theta_1)^2 + \rho_2 (\nabla \theta_2)^2 \right] 
+ \frac{1}{2} \left[ \frac{(\nabla \rho_1)^2}{4\rho_1} + \frac{(\nabla \rho_2)^2}{4\rho_2} \right]+  
\\&+ \frac{\alpha_1}{2} (\rho_1^2 + \rho_2^2) + \alpha_2 \rho_1 \rho_2 -\frac{\Delta}{2} (\rho_1 - \rho_2) -2 \delta \sqrt{\rho_1\rho_2} \cos(\theta_1-\theta_2).
\end{align*}

To simplify the analysis of coupled dynamics and reveal the physical content more transparently, we now introduce a new set of variables corresponding to \textit{total (charge-like)} and \textit{relative (spin-like)} degrees of freedom:
\begin{equation*}
\begin{cases}
     \theta=\theta_1-\theta_2, \\
     \Theta=\frac{1}{2}(\theta_1+\theta_2), \\
    \Pi=\rho_1+\rho_2, \\
     \mathcal{Z}=\frac{1}{2}(\rho_1-\rho_2).
\end{cases}
\end{equation*}

In terms of these new fields, the Lagrangian becomes:
\begin{align*}
\mathcal{L} = & \, \Pi \partial_t \Theta + \mathcal{Z} \partial_t \theta - \frac{1}{2} [\Pi (\nabla \Theta)^2 +\Pi (\frac{\nabla \theta}{2})^2] - \mathcal{Z} \nabla \Theta \nabla \theta  - \frac{1}{8} \left[
\frac{ \left( \frac{1}{2} \nabla \Pi + \nabla \mathcal{Z} \right)^2 }{ \frac{1}{2} \Pi + \mathcal{Z} }
+
\frac{ \left( \frac{1}{2} \nabla \Pi - \nabla \mathcal{Z} \right)^2 }{ \frac{1}{2} \Pi - \mathcal{Z} }
\right] -
\\&-\frac{\alpha_1+\alpha_2}{4} \Pi^2-(\alpha_1 -\alpha_2)\mathcal{Z}^2  + \Delta \mathcal{Z}+\delta\sqrt{\Pi^2-4\mathcal{Z}^2}\cos \theta.
\end{align*}

The new canonically conjugate pairs are:
\begin{equation*}
\begin{cases}
    \pi_\theta=\frac{\partial\mathscr{L}}{\partial(\partial_t\theta)}=\mathcal{Z}, \\
    \pi_\Theta=\frac{\partial\mathscr{L}}{\partial(\partial_t\Theta)}=\Pi.
     \end{cases}
\end{equation*}

Then the corresponding Hamiltonian density becomes:
\begin{align*}
    \mathscr{H} &= \pi_\theta \partial_t \theta + \pi_\Theta \partial_t \Theta - \mathscr{L} = \frac{1}{2} \left[ \Pi (\nabla \Theta)^2 + \Pi \left( \frac{\nabla \theta}{2} \right)^2 \right] 
    + \mathcal{Z} \nabla \Theta \nabla \theta + \\
    &\quad + \frac{1}{8} \left[ \frac{\left( \frac{1}{2} \nabla \Pi + \nabla \mathcal{Z} \right)^2}{\frac{1}{2} \Pi + \mathcal{Z}} + \frac{\left( \frac{1}{2} \nabla \Pi - \nabla \mathcal{Z} \right)^2}{\frac{1}{2} \Pi - \mathcal{Z}} \right]  +\frac{\alpha_1 + \alpha_2}{4} \Pi^2 + (\alpha_1 - \alpha_2) \mathcal{Z}^2 - \Delta \mathcal{Z}-\delta\sqrt{\Pi^2-4\mathcal{Z}^2}\cos \theta.
\end{align*}

The dynamical equations for the field variables can be expressed in canonical Hamiltonian form as:
\begin{align*}
\begin{cases}    &\partial_t \pi_\Theta=\partial_t \Pi=-\frac{\delta\mathscr{H}}{\delta\Theta},\\
\\
&\partial_t \pi_\theta=\partial_t \mathcal{Z}=-\frac{\delta\mathscr{H}}{\delta\theta},\\ 
\\
&\partial_t \Theta=\frac{\delta\mathscr{H}}{\delta\pi_\Theta}=\frac{\delta\mathscr{H}}{\delta\Pi},\\
\\
&\partial_t \theta=\frac{\delta\mathscr{H}}{\delta\pi_\theta}=\frac{\delta\mathscr{H}}{\delta\mathcal{Z}}.
\end{cases}
\end{align*}

After calculating the corresponding functional derivatives for each equation, we obtain the following:
\begin{align*}
\begin{cases}    &\partial_t \Pi=\nabla(\Pi\nabla \Theta)+\nabla(\mathcal{Z} \nabla \theta),\\
\\
&\partial_t \mathcal{Z}=\frac{1}{4}\nabla(\Pi\nabla \theta)+\nabla(\mathcal{Z} \nabla \Theta)-\delta \sqrt{\Pi^2-4 \mathcal{Z}^2}\sin \theta ,\\
\\
&\partial_t \Theta= \frac{1}{2} \left[ (\nabla \Theta)^2 +  \left( \frac{\nabla \theta}{2} \right)^2 \right] - \frac{1}{8} \left[ \frac{1}{2} \frac{\left( \frac{1}{2} \nabla \Pi + \nabla \mathcal{Z} \right)^2}{(\frac{1}{2} \Pi + \mathcal{Z})^2} + \frac{1}{2} \frac{\left( \frac{1}{2} \nabla \Pi - \nabla \mathcal{Z} \right)^2}{(\frac{1}{2} \Pi - \mathcal{Z})^2} \right] + \frac{\alpha_1+\alpha_2}{2}{\Pi} - \frac{1}{8} \nabla  \left[
\frac{ \frac{1}{2} \nabla \Pi + \nabla \mathcal{Z}  }{ \frac{1}{2} \Pi + \mathcal{Z} }
+
\frac{ \frac{1}{2} \nabla \Pi - \nabla \mathcal{Z}  }{ \frac{1}{2} \Pi - \mathcal{Z} }
\right] -\frac{\delta \Pi}{\sqrt{\Pi^2-4\mathcal{Z}^2}}  \cos \theta,\\ \\
&\partial_t \theta= \nabla \Theta \nabla \theta +2(\alpha_1-\alpha_2)\mathcal{Z}-\Delta - \frac{1}{8} \left[  \frac{\left( \frac{1}{2} \nabla \Pi + \nabla \mathcal{Z} \right)^2}{(\frac{1}{2} \Pi + \mathcal{Z})^2} - \frac{\left( \frac{1}{2} \nabla \Pi - \nabla \mathcal{Z} \right)^2}{(\frac{1}{2} \Pi - \mathcal{Z})^2} \right] -\frac{1}{8} \nabla  \left[
2 \frac{ \frac{1}{2} \nabla \Pi + \nabla \mathcal{Z}  }{ \frac{1}{2} \Pi + \mathcal{Z} }
-2
\frac{ \frac{1}{2} \nabla \Pi - \nabla \mathcal{Z}  }{ \frac{1}{2} \Pi - \mathcal{Z} }
\right]+\frac{4 \delta \mathcal{Z}}{\sqrt{\Pi^2-4\mathcal{Z}^2}} \cos \theta.
\end{cases}
\end{align*}

To obtain a more realistic description of the system dynamics, it is essential to account for relaxation effects, which govern the evolution of spin degrees of freedom. In our approach, relaxation is incorporated phenomenologically by adding gradient-dependent dissipative terms proportional to the variational derivatives of the Hamiltonian. This form ensures that the relaxation drives the system toward energy minimization. Specifically, we modify the equations for the spin-phase variables $\Theta$ and $\theta$ as follows:
\begin{align*}
\begin{cases}    
&\partial_t \Theta=\frac{\delta\mathscr{H}}{\delta\Pi} - \gamma_\Theta\frac{\delta\mathscr{H}}{\delta \Theta}=\frac{\delta\mathscr{H}}{\delta\Pi} + \gamma_\Theta[\nabla(\Pi\nabla \Theta)+\nabla(\mathcal{Z} \nabla \theta)],\\
\\
&\partial_t \theta=\frac{\delta\mathscr{H}}{\delta\mathcal{Z}}-  \gamma_\theta\frac{\delta\mathscr{H}}{\delta \theta} =\frac{\delta\mathscr{H}}{\delta\mathcal{Z}}+  \gamma_\theta[\frac{1}{4}\nabla(\Pi\nabla \theta)+\nabla(\mathcal{Z} \nabla \Theta)-\delta \sqrt{\Pi^2-4 \mathcal{Z}^2}\sin \theta].
\end{cases}
\end{align*}

These additional terms act to suppress spatial inhomogeneities in the corresponding spin-phase fields and thus model pure spin relaxation in the hydrodynamic framework, while preserving the overall structure of the canonical field equations.

Finally, these equations will take the form
\begin{align*}
\begin{cases}    
&\partial_t \Theta= \frac{1}{2} \left[ (\nabla \Theta)^2 +  \left( \frac{\nabla \theta}{2} \right)^2 \right] - \frac{1}{8} \left[ \frac{1}{2} \frac{\left( \frac{1}{2} \nabla \Pi + \nabla \mathcal{Z} \right)^2}{(\frac{1}{2} \Pi + \mathcal{Z})^2} + \frac{1}{2} \frac{\left( \frac{1}{2} \nabla \Pi - \nabla \mathcal{Z} \right)^2}{(\frac{1}{2} \Pi - \mathcal{Z})^2} \right] + \frac{\alpha_1+\alpha_2}{2}{\Pi} - \frac{1}{8} \nabla  \left[
\frac{ \frac{1}{2} \nabla \Pi + \nabla \mathcal{Z}  }{ \frac{1}{2} \Pi + \mathcal{Z} }
+
\frac{ \frac{1}{2} \nabla \Pi - \nabla \mathcal{Z}  }{ \frac{1}{2} \Pi - \mathcal{Z} }
\right] -
\\&\quad \quad \quad \quad \quad \quad \quad \quad \quad \quad \quad \quad \quad \quad \quad \quad \quad \quad \quad \quad \quad \quad \quad \quad  \quad \quad \quad \quad \quad \quad \quad -\frac{\delta \Pi}{\sqrt{\Pi^2-4\mathcal{Z}^2}}  \cos \theta+\gamma_\Theta[\nabla(\Pi\nabla \Theta)+\nabla(\mathcal{Z} \nabla \theta)]  ,\\ \\
&\partial_t \theta= \nabla \Theta \nabla \theta +2(\alpha_1-\alpha_2)\mathcal{Z}-\Delta - \frac{1}{8} \left[  \frac{\left( \frac{1}{2} \nabla \Pi + \nabla \mathcal{Z} \right)^2}{(\frac{1}{2} \Pi + \mathcal{Z})^2} -  \frac{\left( \frac{1}{2} \nabla \Pi - \nabla \mathcal{Z} \right)^2}{(\frac{1}{2} \Pi - \mathcal{Z})^2} \right] -\frac{1}{8} \nabla  \left[
2 \frac{ \frac{1}{2} \nabla \Pi + \nabla \mathcal{Z}  }{ \frac{1}{2} \Pi + \mathcal{Z} }
-2
\frac{ \frac{1}{2} \nabla \Pi - \nabla \mathcal{Z}  }{ \frac{1}{2} \Pi - \mathcal{Z} }
\right] + \\
&\quad \quad \quad \quad \quad \quad \quad \quad \quad \quad \quad \quad \quad \quad \quad  \quad \quad \quad \quad \quad \quad \quad +\frac{4 \delta \mathcal{Z}}{\sqrt{\Pi^2-4\mathcal{Z}^2}} \cos \theta+\gamma_\theta [\frac{1}{4}\nabla(\Pi\nabla \theta)+\nabla(\mathcal{Z} \nabla \Theta)-\delta \sqrt{\Pi^2-4 \mathcal{Z}^2}\sin \theta].
\end{cases}
\end{align*}

Next, we consider relaxation in the equations for $\Pi$ and $\mathcal{Z}$. We do not include relaxation in the equation for the total density $\Pi = \rho_1 + \rho_2$, since it represents a conserved quantity in the absence of pumping or decay. Introducing dissipation here would violate particle number conservation. In contrast, spin imbalance $\mathcal{Z} = \frac{1}{2}(\rho_1 - \rho_2)$ is not conserved and can relax. However, direct inclusion of relaxation into its equation may lead to unphysical values where $|\mathcal{Z}| > \frac{1}{2} \Pi$. To avoid this, we parametrize $\mathcal{Z}$ as:
\begin{equation*}
     \mathcal{Z}=\frac{1}{2}(\rho_1-\rho_2)=\frac{1}{2} \Pi \cos \Omega,
\end{equation*}
which ensures that $\mathcal{Z}$ remains within physical bounds. This substitution allows us to introduce relaxation via a damping term in the equation for $\Omega$, ensuring both mathematical regularity and physical consistency. So, let's rewrite the equation for $\mathcal{Z}$ in terms of $\Pi$ and $\Omega$:
\begin{align*} 
    &\partial_t \mathcal{Z}=\frac{1}{2} \partial_t\Pi \cos \Omega-\frac{1}{2} \Pi \sin\Omega \partial_t \Omega =-\frac{\delta\mathscr{H}}{\delta\theta}= \frac{1}{4}\nabla(\Pi\nabla \theta)+\nabla(\frac{1}{2} \Pi \cos \Omega \nabla \Theta) -\delta \Pi \sin \Omega \sin \theta.
\end{align*}

Let's simplify and rewrite it as follows:
\begin{align*} 
    &\partial_t \Omega = \nabla \Omega  \nabla \Theta - \frac{\nabla  \Pi \nabla \theta}{2 \Pi} \sin \Omega -\frac{\nabla^2 \theta}{2} \sin \Omega-\frac{1}{2} \cos \Omega \nabla \Omega \nabla \theta +2 \delta \sin \theta.
\end{align*}

And only now we will introduce the relaxation term in this equation:
\begin{align*} 
    &\partial_t \Omega = \nabla \Omega  \nabla \Theta - \frac{\nabla  \Pi \nabla \theta}{2 \Pi} \sin \Omega -\frac{\nabla^2 \theta}{2} \sin \Omega-\frac{1}{2} \cos \Omega \nabla \Omega \nabla \theta +2 \delta \sin \theta -\gamma_\Omega \frac{\delta \mathscr{H}}{\delta \Omega}.
\end{align*}

Let us write the Hamiltonian as a function of the fields $\Theta, \ \theta, \ \Pi$ and $\Omega$:
\begin{align*}
    \mathscr{H} & = \frac{1}{2} \left[ \Pi (\nabla \Theta)^2 + \Pi \left( \frac{\nabla \theta}{2} \right)^2 \right] 
    + \frac{1}{2}\Pi \cos \Omega \nabla \Theta \nabla \theta + \frac{1}{8} \left[ \frac{\left( \frac{1}{2} \nabla \Pi + \frac{1}{2} \nabla (\Pi \cos \Omega) \right)^2}{\frac{1}{2} \Pi(1 + \cos \Omega)} + \frac{\left( \frac{1}{2} \nabla \Pi - \frac{1}{2}\nabla (\Pi \cos \Omega) \right)^2}{\frac{1}{2} \Pi(1 - \cos \Omega)} \right]  +\\
    &\quad \quad \quad \quad \quad \quad \quad \quad \quad \quad \quad \quad \quad \quad \quad \quad \quad \quad \quad \quad \quad \quad \quad  +\frac{\alpha_1 + \alpha_2}{4} \Pi^2 + \frac{\alpha_1 - \alpha_2}{4} (\Pi \cos \Omega)^2 -  \frac{1}{2}  \Pi \Delta \cos \Omega -\delta \Pi \sin \Omega \cos \theta.
\end{align*}

By simplifying the expression, we obtain the more compact form:
\begin{align*}
    \mathscr{H} & = \frac{1}{2} \left[ \Pi (\nabla \Theta)^2 + \Pi \left( \frac{\nabla \theta}{2} \right)^2 \right] 
    + \frac{1}{2}\Pi \cos \Omega \nabla \Theta \nabla \theta  + \frac{1}{8} \left[ \frac{(\nabla \Pi)^2}{\Pi} +\Pi(\nabla \Omega)^2\right]  + \\
    & \quad \quad \quad \quad \quad \quad \quad \quad \quad \quad \quad \quad \quad \quad \quad \quad \quad \quad \quad \quad +\frac{\alpha_1 + \alpha_2}{4} \Pi^2 + \frac{\alpha_1 - \alpha_2}{4} (\Pi \cos \Omega)^2 -  \frac{1}{2}  \Pi \Delta \cos \Omega  -\delta \Pi \sin \Omega \cos \theta.
\end{align*}

The next step involves computing the variational derivative of the functional \(\mathscr{H}\) with respect to \(\Omega\):
\begin{align*}
    &\frac{\delta \mathscr{H}}{\delta \Omega} =
- \frac{1}{2} \Pi \sin\Omega\, \nabla \Theta  \nabla \theta
- \frac{1}{2} (\alpha_1 - \alpha_2) \Pi^2 \cos\Omega \sin\Omega
+\frac{1}{2} \Pi \Delta \sin\Omega - \delta \Pi \cos \Omega \cos \theta
- \frac{1}{4}\nabla (\Pi \nabla \Omega).
\end{align*}
This term will be used to define dissipative dynamics in the evolution of  $\Omega$.

Finally, taking into account the relaxation terms, the system of equations of motion takes the following form:
\begin{align*}
\begin{cases}    
&\partial_t \Theta= \frac{1}{2} \left[ (\nabla \Theta)^2 +  \left( \frac{\nabla \theta}{2} \right)^2 \right] - \frac{1}{8} \left[ \frac{1}{2} \frac{\left( \frac{1}{2} \nabla \Pi + \nabla \mathcal{Z} \right)^2}{(\frac{1}{2} \Pi + \mathcal{Z})^2} + \frac{1}{2} \frac{\left( \frac{1}{2} \nabla \Pi - \nabla \mathcal{Z} \right)^2}{(\frac{1}{2} \Pi - \mathcal{Z})^2} \right] + \frac{\alpha_1+\alpha_2}{2}{\Pi} - \frac{1}{8} \nabla  \left[
\frac{ \frac{1}{2} \nabla \Pi + \nabla \mathcal{Z}  }{ \frac{1}{2} \Pi + \mathcal{Z} }
+
\frac{ \frac{1}{2} \nabla \Pi - \nabla \mathcal{Z}  }{ \frac{1}{2} \Pi - \mathcal{Z} }
\right] - 
\\&\quad \quad \quad \quad \quad \quad \quad \quad \quad \quad \quad \quad \quad \quad \quad \quad \quad \quad \quad \quad \quad \quad \quad \quad  \quad \quad \quad \quad \quad \quad \quad -\frac{\delta \Pi}{\sqrt{\Pi^2-4\mathcal{Z}^2}}  \cos \theta+\gamma_\Theta[\nabla(\Pi\nabla \Theta)+\nabla(\mathcal{Z} \nabla \theta)]  ,\\ \\
&\partial_t \theta= \nabla \Theta \nabla \theta +2(\alpha_1-\alpha_2)\mathcal{Z}-\Delta - \frac{1}{8} \left[  \frac{\left( \frac{1}{2} \nabla \Pi + \nabla \mathcal{Z} \right)^2}{(\frac{1}{2} \Pi + \mathcal{Z})^2} -  \frac{\left( \frac{1}{2} \nabla \Pi - \nabla \mathcal{Z} \right)^2}{(\frac{1}{2} \Pi - \mathcal{Z})^2} \right] -\frac{1}{8} \nabla  \left[
2 \frac{ \frac{1}{2} \nabla \Pi + \nabla \mathcal{Z}  }{ \frac{1}{2} \Pi + \mathcal{Z} }
-2
\frac{ \frac{1}{2} \nabla \Pi - \nabla \mathcal{Z}  }{ \frac{1}{2} \Pi - \mathcal{Z} }
\right] + \\
&\quad \quad \quad \quad \quad \quad \quad \quad \quad \quad \quad \quad \quad \quad \quad  \quad \quad \quad \quad \quad \quad \quad +\frac{4 \delta \mathcal{Z}}{\sqrt{\Pi^2-4\mathcal{Z}^2}} \cos \theta+\gamma_\theta [\frac{1}{4}\nabla(\Pi\nabla \theta)+\nabla(\mathcal{Z} \nabla \Theta)-\delta \sqrt{\Pi^2-4 \mathcal{Z}^2}\sin \theta] \\
\\
&\partial_t \Pi=\nabla(\Pi\nabla \Theta)+\nabla(\mathcal{Z} \nabla \theta),\\
\\
&\partial_t \Omega = \nabla \Omega  \nabla \Theta - \frac{\nabla  \Pi \nabla \theta}{2 \Pi} \sin \Omega -\frac{\nabla^2 \theta}{2} \sin \Omega-\frac{1}{2} \cos \Omega \nabla \Omega \nabla \theta +2 \delta \sin \theta - \\
&\quad \quad \quad \quad \quad \quad \quad  \quad \quad \quad \quad \quad \quad \quad  -\gamma_\Omega[- \frac{1}{2} \Pi \sin\Omega\, \nabla \Theta  \nabla \theta
- \frac{1}{2} (\alpha_1 - \alpha_2) \Pi^2 \cos\Omega \sin\Omega
+ \frac{1}{2} \Pi \Delta \sin\Omega - \delta \Pi \cos \Omega \cos \theta
- \frac{1}{4}\nabla (\Pi \nabla \Omega) ].
\end{cases}
\end{align*}

By substituting $\mathcal{Z} = \frac{1}{2} \Pi \cos \Omega$ into the equations above and simplifying, the system becomes:
\begin{align}
\begin{cases}    \label{systemwithrelax0}
&\partial_t \Theta= \frac{1}{2} \left[ (\nabla \Theta)^2 +  \left( \frac{\nabla \theta}{2} \right)^2 \right] - \frac{1}{8} \left[ (\frac{\nabla \Pi}{\Pi}+ \nabla \Omega \cot{\Omega})^2 +\frac{(\nabla \Omega)^2}{\sin^2 \Omega} \right] + \\
&\qquad\qquad\qquad\qquad\qquad\qquad\qquad + \frac{\alpha_1+\alpha_2}{2}{\Pi} - \frac{1}{4} \nabla  \left[ \frac{\nabla \Pi}{\Pi}+\nabla \Omega \cot{\Omega} \right]  -\frac{\delta \cos \theta }{\sin \Omega} +\gamma_\Theta[\nabla(\Pi\nabla \Theta)+\frac{1}{2}\nabla(\Pi \cos \Omega \nabla \theta)]  ,\\ \\
&\partial_t \theta= \nabla \Theta \nabla \theta +(\alpha_1-\alpha_2)\Pi \cos \Omega-\Delta + \frac{1}{2} \frac{\nabla \Omega}{\sin \Omega} \left[ \frac{\nabla \Pi }{\Pi}+\nabla \Omega \cot{\Omega}  \right]  + \\
&\qquad\qquad\qquad\qquad\qquad\qquad\qquad\qquad \qquad\qquad\qquad +\frac{1}{2} \nabla  (\frac{\nabla \Omega}{\sin \Omega}) + 2 \delta \cot \Omega \cos \theta+ \gamma_\theta [\frac{1}{4}\nabla(\Pi\nabla \theta)+\frac{1}{2}\nabla(\Pi \cos \Omega \nabla \Theta)-\delta \Pi \sin \Omega \sin \theta] , \\
\\
&\partial_t \Pi=\nabla(\Pi\nabla \Theta)+\frac{1}{2}\nabla(\Pi \cos \Omega \nabla \theta),\\
\\
&\partial_t \Omega = \nabla \Omega  \nabla \Theta - \frac{\nabla  \Pi \nabla \theta}{2 \Pi} \sin \Omega -\frac{\nabla^2 \theta}{2} \sin \Omega-\frac{1}{2} \cos \Omega \nabla \Omega \nabla \theta +2 \delta \sin \theta + \\
&\quad \quad \quad \quad \quad \quad \quad  \quad \quad \quad \quad \quad \quad \quad  +\gamma_\Omega[\ \frac{1}{2} \Pi \sin\Omega\, \nabla \Theta  \nabla \theta
+ \frac{1}{2} (\alpha_1 - \alpha_2) \Pi^2 \cos\Omega \sin\Omega
- \frac{1}{2} \Pi \Delta \sin\Omega + \delta \Pi \cos \Omega \cos \theta
+ \frac{1}{4}\nabla (\Pi \nabla \Omega) ].
\end{cases}
\end{align}

The system \ref{systemwithrelax0} describes the dynamics of a spinor polariton condensate in terms of the global phase~$\Theta$, total density~$\Pi$, relative phase~$\theta$, and polarization angle~$\Omega$. It incorporates nonlinear interactions, dissipation, and spin-dependent effects.

In the case of a spatially uniform condensate, where all gradients vanish $\nabla \Theta=\nabla \theta = \nabla \Pi= \nabla \Omega=0$, the system reduces to a set of ordinary differential equations:
\begin{align*}
\begin{cases}    
&\partial_t \Theta=\frac{\alpha_1+\alpha_2}{2}{\Pi}-\frac{\delta \cos \theta}{\sin \Omega},\\ \\
&\partial_t \theta=  (\alpha_1-\alpha_2)\Pi \cos \Omega-\Delta+2\delta \cot \Omega \cos \theta -\gamma_\theta \delta \Pi \sin \Omega \sin \theta  ,\\
\\
&\partial_t \Pi=0,\\
\\
&\partial_t \Omega =2\delta \sin \theta+\gamma_\Omega [
\frac{1}{2} (\alpha_1 - \alpha_2) \Pi^2 \cos\Omega \sin\Omega
- \frac{1}{2} \Pi \Delta \sin\Omega+\delta \Pi \cos \Omega \cos \theta 
].
\end{cases}
\end{align*}

\section*{Supplemental Note 2: Representation via Stokes vector}

To provide an intuitive interpretation, we introduce the Stokes (or polarization) vector $\mathbf{S} = (S_x, S_y, S_z)$ defined by:
\begin{align*}
\begin{cases}
S_x &= \Pi \sin \Omega \cos \theta, \\
S_y &= \Pi \sin \Omega \sin \theta, \\
S_z &= \Pi \cos \Omega.
\end{cases}
\end{align*}

The inverse transformation reads:
\begin{align*}
\Pi &= \sqrt{S_x^2 + S_y^2 + S_z^2}, \\
\theta &= \arg(S_x + i S_y) \Rightarrow \theta=\arctan{\frac{S_y}{S_x}}, \\
\cos \Omega &= \frac{S_z}{\Pi}, \quad \sin \Omega = \frac{\sqrt{S_x^2 + S_y^2}}{\Pi}.
\end{align*}

The next step is reformulating the dynamics in terms of $(\Theta, \Pi, \mathbf{S})$. Since the general case is cumbersome, we restrict our consideration to the homogeneous condensate ($\nabla \Theta=\nabla \theta = \nabla \Pi= \nabla \Omega=0$):
\begin{align*}
\begin{cases}    
&\partial_t \Theta= \frac{\alpha_1+\alpha_2}{2}{\Pi}- \frac{\delta \Pi S_x}{S_\perp^2},\\ \\
&\partial_t \Pi=0,\\
\\
&\partial_t S_x =\frac{S_z S_x}{S_\perp}\partial_t \Omega-S_y \partial_t \theta
\\ \\
&\partial_t S_y= \frac{S_z S_y}{S_\perp}\partial_t \Omega+S_x \partial_t \theta
\\ \\
&\partial_t S_z =-S_\perp \partial_t \Omega
\end{cases}
\end{align*}
where $S_\perp = \sqrt{S_x^2 + S_y^2}$ is the in-plane pseudospin projection. The derivatives $\partial_t \theta$ and $\partial_t \Omega$ are given by

\begin{align*}
\begin{cases}    
&\partial_t \theta= (\alpha_1-\alpha_2)S_z-\Delta+2\delta \frac{S_z S_x}{S_\perp^2} -\gamma_\theta \delta S_y \\
\\
&\partial_t \Omega = 2\delta \frac{S_y}{S_\perp}+\gamma_\Omega [
 \frac{(\alpha_1 - \alpha_2)}{2}  S_z S_\perp
- \frac{\Delta}{2}  S_\perp+\delta \frac{S_z S_x}{S_\perp}
].
\end{cases}
\end{align*}

The time evolution of the Stokes vector can thus be written compactly as:
\begin{align} \label{eqforStokesvector}
\partial_t \mathbf{S} = 
\mathbf{B}_\text{eff}(\mathbf{S}) \times \mathbf{S} + \boldsymbol{\Gamma}(\mathbf{S}),
\end{align}
where
\[
\mathbf{B}_\text{eff}(\mathbf{S}) = 
\begin{pmatrix}
-2\delta \\
0 \\
-\Delta + (\alpha_1 - \alpha_2) S_z
\end{pmatrix},
\quad
\boldsymbol{\Gamma}(\mathbf{S}) = 
\begin{pmatrix}
\gamma_\Omega S_x S_z \left( \frac{(\alpha_1 - \alpha_2)}{2} S_z - \frac{\Delta}{2} + \delta \frac{S_z S_x}{S_\perp^2} \right) + \gamma_\theta \delta S_y^2 \\
\gamma_\Omega S_y S_z \left( \frac{(\alpha_1 - \alpha_2)}{2} S_z - \frac{\Delta}{2} + \delta \frac{S_z S_x}{S_\perp^2} \right) - \gamma_\theta \delta S_x S_y \\
-\gamma_\Omega S_\perp^2 \left( \frac{(\alpha_1 - \alpha_2)}{2} S_z - \frac{\Delta}{2} + \delta \frac{S_z S_x}{S_\perp^2} \right)
\end{pmatrix}.
\]

This form highlights the spin-precession in the effective nonlinear magnetic field $\mathbf{B}_\text{eff}$ along with anisotropic dissipation encoded in $\boldsymbol{\Gamma}(\mathbf{S})$. Equation~\eqref{eqforStokesvector} can also be rewritten as a generalized Landau–Lifshitz–Gilbert equation:
\begin{align} \label{LLG}
\partial_t \mathbf{S} = 
\mathbf{B}_\text{eff} (\mathbf{S}) \times \mathbf{S} 
- \lambda \mathbf{S} \times  \left[\mathbf{B}_\text{eff} (\mathbf{S}) \times \mathbf{S}\right] + \mathbf{F}_\text{aniso}(\mathbf{S}),
\end{align}
where $\lambda = \gamma_\Omega / 2$ is the damping coefficient, and the anisotropic force is given by:
\[
\mathbf{F}_\text{aniso}(\mathbf{S}) = \delta \left(\gamma_\theta - \gamma_\Omega \frac{S^2}{S_\perp^2} \right)
\begin{pmatrix}
S_y^2 \\
- S_x S_y \\
0
\end{pmatrix}.
\]
This term reflects an intrinsic anisotropy in the dissipative response of the system, which can strongly influence the polarization dynamics.

Thus, the generalized LLG equation~\eqref{LLG} describes the key physical ingredients of driven-dissipative spinor systems: coherent precession, nonlinear and spatially dependent damping, and anisotropic relaxation—going beyond the standard LLG framework and providing insight into the rich dynamical behavior of polariton condensates.

\section*{Supplemental Note 3: Dispersion of elementary excitations}

In this section, we derive the dispersion relation for elementary excitations of the condensate. For simplicity, we begin with the conservative case, i.e., when there is no relaxation in the system:
\begin{align} 
\begin{cases}    \label{finalsystemwithoutrelax1}
&\partial_t \Theta= \frac{1}{2} \left[ (\nabla \Theta)^2 +  \left( \frac{\nabla \theta}{2} \right)^2 \right] - \frac{1}{8} \left[ (\frac{\nabla \Pi}{\Pi}+ \nabla \Omega \cot{\Omega})^2 +\frac{(\nabla \Omega)^2}{\sin^2 \Omega} \right] + \frac{\alpha_1+\alpha_2}{2}{\Pi} - \frac{1}{4} \nabla  \left[ \frac{\nabla \Pi}{\Pi}+\nabla \Omega \cot{\Omega} \right]  -\frac{\delta \cos \theta }{\sin \Omega}  ,\\ \\
&\partial_t \theta= \nabla \Theta \nabla \theta +(\alpha_1-\alpha_2)\Pi \cos \Omega-\Delta + \frac{1}{2} \frac{\nabla \Omega}{\sin \Omega} \left[ \frac{\nabla \Pi }{\Pi}+\nabla \Omega \cot{\Omega}  \right]  +\frac{1}{2} \nabla  (\frac{\nabla \Omega}{\sin \Omega}) + 2 \delta \cot \Omega \cos \theta, \\
\\
&\partial_t \Pi=\nabla(\Pi\nabla \Theta)+\frac{1}{2}\nabla(\Pi \cos \Omega \nabla \theta),\\
\\
&\partial_t \Omega = \nabla \Omega  \nabla \Theta - \frac{\nabla  \Pi \nabla \theta}{2 \Pi} \sin \Omega -\frac{\nabla^2 \theta}{2} \sin \Omega-\frac{1}{2} \cos \Omega \nabla \Omega \nabla \theta +2 \delta \sin \theta.
\end{cases}
\end{align}

We seek a stationary and spatially homogeneous condensate state that minimizes 
the system’s energy. The corresponding equilibrium conditions for the fields 
\begin{equation*}
\begin{cases}
    \partial_t \Pi = 0,\\
    \partial_t \Omega = 0,\\
    \partial_t \theta = 0,\\
    \partial_t \Theta = \mu,
\end{cases}
\quad \text{evaluated at } (\Pi_0, \Theta_0, \theta_0, \Omega_0).
\end{equation*}
This leads to:
\begin{align*}
\begin{cases}    
&\partial_t \Theta_0= \frac{\alpha_1+\alpha_2}{2}{\Pi_0}-\frac{\delta \cos \theta_0}{\sin \Omega_0}=\mu,\\ \\
&\partial_t \theta_0=  (\alpha_1-\alpha_2)\Pi_0 \cos \Omega_0-\Delta+2\delta \cot \Omega_0 \cos \theta_0  =0 ,\\
\\
&\partial_t \Pi_0=0,\\
\\
&\partial_t \Omega_0   = 2\delta \sin \theta_0=0.
\end{cases}
\end{align*}

The condition \(\partial_t \Omega_0 = 0\) implies \(\sin \theta_0 = 0\), and thus \(\theta_0 = 0\). Substituting this into the remaining equilibrium equations yields the constraints:
\[
(\alpha_1 - \alpha_2) \Pi_0 \cos \Omega_0 - \Delta + 2\delta \cot \Omega_0 = 0,
\]
and for the chemical potential
\[
\mu = \frac{\alpha_1 + \alpha_2}{2} \Pi_0 - \frac{\delta}{\sin \Omega_0}.
\]

To analyze the stability of the equilibrium state, we consider small perturbations around the stationary homogeneous background:
\begin{equation}
\begin{cases} \label{ansatz1} 
    \Pi=\Pi_0+\updelta\Pi(\bm{r},t), \\
    \Theta=\mu t+\updelta\Theta(\bm{r},t), \\
    \theta=\theta_0+\updelta\theta(\bm{r},t), \\
    \Omega=\Omega_0+\updelta\Omega(\bm{r},t),
\end{cases}
\end{equation}
where $|\delta f| \ll |f_0|$ for $f=\Pi, \Theta, \theta \ \text{or} \ \Omega$. 
Substituting this ansatz \ref{ansatz1} into the system \ref{finalsystemwithoutrelax1}, linearizing with respect to $\delta \Omega, \delta \Pi, \delta \Theta, \delta \theta$ and using $\theta_0=0$, we obtain the following system:
\begin{align*}
\begin{cases}   
&\partial_t \updelta \Theta=   \frac{\alpha_1+\alpha_2}{2}\updelta\Pi - \frac{1}{4} \frac{\nabla^2 \updelta \Pi}{\Pi_0}- \frac{1}{4} \cot{\Omega_0} \nabla^2 \updelta\Omega + \frac{\delta}{\sin \Omega_0}\cot\Omega_0 \updelta \Omega ,\\ \\
&\partial_t \updelta \theta= - (\alpha_1-\alpha_2)[\Pi_0 \sin\Omega_0\updelta\Omega-\cos\Omega_0\updelta\Pi] +\frac{1}{2} \frac{\nabla^2 \updelta \Omega}{\sin \Omega_0} - \frac{2 \delta}{\sin^2\Omega_0}\updelta \Omega  , \\
\\
&\partial_t \updelta\Pi= \Pi_0 \nabla^2 \updelta \Theta+\frac{1}{2} \Pi_0 \cos \Omega_0 \nabla^2 \updelta \theta,\\
\\
&\partial_t \updelta \Omega = -\frac{\nabla^2 \updelta \theta}{2} \sin \Omega_0+2 \delta \updelta \theta.
\end{cases}
\end{align*}

Having linearized the system around the stationary homogeneous state, we proceed to study the dynamics of small perturbations. To do so, we represent all deviations as superpositions of real harmonic modes:
\begin{equation} \label{harmonics}
\updelta f(\mathbf{r}, t) = A_f \cos(\mathbf{k} \cdot \mathbf{r} - \omega t) + B_f \sin(\mathbf{k} \cdot \mathbf{r} - \omega t),
\quad f \in \{ \Theta, \theta, \Pi, \Omega \}
\end{equation}

Substituting this ansatz into the linearized equations and using the identities for derivatives:
\begin{align} \label{identities}
\partial_t \delta f &= \omega \left[ A_f \sin(\mathbf{k} \cdot \mathbf{r} - \omega t) - B_f \cos(\mathbf{k} \cdot \mathbf{r} - \omega t) \right], \\ \notag
\nabla \delta f &= -\mathbf{k} \left[ A_f \sin(\mathbf{k} \cdot \mathbf{r} - \omega t) - B_f \cos(\mathbf{k} \cdot \mathbf{r} - \omega t) \right], \\ \notag
\nabla^2 \delta f &= -k^2 \delta f, \notag
\end{align}
we rewrite the full system in harmonic form:
\begin{align*}
\begin{cases}   
&\omega \left[ A_\Theta \sin(\mathbf{k} \cdot \mathbf{r} - \omega t) - B_\Theta \cos(\mathbf{k} \cdot \mathbf{r} - \omega t) \right]=  \frac{\alpha_1+\alpha_2}{2}[A_\Pi \cos(\mathbf{k} \cdot \mathbf{r} - \omega t) + B_\Pi \sin(\mathbf{k} \cdot \mathbf{r} - \omega t)]+ k^2 \frac{1}{4 \Pi_0}  [A_\Pi \cos(\mathbf{k} \cdot \mathbf{r} - \omega t)+ \\
&  + B_\Pi \sin(\mathbf{k} \cdot \mathbf{r} - \omega t)]+\frac{k^2}{4} \cot{\Omega_0} [A_\Omega \cos(\mathbf{k} \cdot \mathbf{r} - \omega t) + B_\Omega \sin(\mathbf{k} \cdot \mathbf{r} - \omega t)] +\frac{\delta}{\sin \Omega_0}\cot\Omega_0 [A_\Omega \cos(\mathbf{k} \cdot \mathbf{r} - \omega t) + \\
&+B_\Omega \sin(\mathbf{k} \cdot \mathbf{r} - \omega t)] ,\\ \\
&\omega \left[ A_\theta \sin(\mathbf{k} \cdot \mathbf{r} - \omega t) - B_\theta \cos(\mathbf{k} \cdot \mathbf{r} - \omega t) \right]=  -(\alpha_1-\alpha_2)[\Pi_0 \sin\Omega_0[A_\Omega \cos(\mathbf{k} \cdot \mathbf{r} - \omega t) + B_\Omega \sin(\mathbf{k} \cdot \mathbf{r} - \omega t)]- \\
&-\cos\Omega_0[A_\Pi \cos(\mathbf{k} \cdot \mathbf{r} - \omega t) +B_\Pi \sin(\mathbf{k} \cdot \mathbf{r} - \omega t)]] -\frac{1}{2} \frac{k^2 }{\sin \Omega_0}[A_\Omega \cos(\mathbf{k} \cdot \mathbf{r} - \omega t) + B_\Omega \sin(\mathbf{k} \cdot \mathbf{r} - \omega t)] - \frac{2 \delta}{\sin^2\Omega_0}[A_\Omega \cos(\mathbf{k} \cdot \mathbf{r} - \omega t) + , \\
&+ B_\Omega \sin(\mathbf{k} \cdot \mathbf{r} - \omega t)] , \\
\\
&\omega \left[ A_\Pi \sin(\mathbf{k} \cdot \mathbf{r} - \omega t) - B_\Pi \cos(\mathbf{k} \cdot \mathbf{r} - \omega t) \right]= -k^2 \Pi_0 [A_\Theta \cos(\mathbf{k} \cdot \mathbf{r} - \omega t) + B_\Theta \sin(\mathbf{k} \cdot \mathbf{r} - \omega t)]-\frac{k^2}{2} \Pi_0 \cos \Omega_0 [A_\theta \cos(\mathbf{k} \cdot \mathbf{r} - \omega t) + \\
&+B_\theta \sin(\mathbf{k} \cdot \mathbf{r} - \omega t)],\\
\\
&\omega \left[ A_\Omega \sin(\mathbf{k} \cdot \mathbf{r} - \omega t) - B_\Omega \cos(\mathbf{k} \cdot \mathbf{r} - \omega t) \right] =\frac{k^2 }{2} \sin \Omega_0[A_\theta \cos(\mathbf{k} \cdot \mathbf{r} - \omega t) + B_\theta \sin(\mathbf{k} \cdot \mathbf{r} - \omega t)]+2 \delta  [A_\theta \cos(\mathbf{k} \cdot \mathbf{r} - \omega t) +\\
&+B_\theta \sin(\mathbf{k} \cdot \mathbf{r} - \omega t)].
\end{cases}
\end{align*}

Next, we express the small deviations as a superposition of real harmonic functions, as defined in Eq.~\eqref{harmonics}. By applying the  identities given in Eq.~\eqref{identities} and substituting this ansatz into the linearized dynamical equations, we derive a set of coupled equations governing the amplitudes. Dividing terms proportional to \(\cos(\mathbf{k} \cdot \mathbf{r} - \omega t)\) and \(\sin(\mathbf{k} \cdot \mathbf{r} - \omega t)\), we obtain a system of eight linear algebraic equations for the unknown amplitudes:
\begin{align*}
\begin{cases}
&-\omega B_\Theta = \frac{\alpha_1 + \alpha_2}{2} A_\Pi + \frac{k^2}{4 \Pi_0} A_\Pi + \frac{k^2}{4} \cot \Omega_0 A_\Omega +\frac{\delta}{\sin \Omega_0} \cot \Omega_0 A_\Omega , \\
&  \omega A_\Theta = \frac{\alpha_1 + \alpha_2}{2} B_\Pi + \frac{k^2}{4 \Pi_0} B_\Pi + \frac{k^2}{4} \cot \Omega_0 B_\Omega + \frac{\delta}{\sin \Omega_0}  \cot \Omega_0 B_\Omega, \\
&-\omega B_\theta = -(\alpha_1 - \alpha_2)\left( \Pi_0 \sin \Omega_0 A_\Omega - \cos \Omega_0 A_\Pi \right) - \frac{k^2}{2 \sin \Omega_0} A_\Omega - \frac{2 \delta}{\sin^2 \Omega_0} A_\Omega ,   \\
& 
\omega A_\theta = -(\alpha_1 - \alpha_2)\left( \Pi_0 \sin \Omega_0 B_\Omega - \cos \Omega_0 B_\Pi \right) - \frac{k^2}{2 \sin \Omega_0} B_\Omega -   \frac{2 \delta}{\sin^2 \Omega_0} B_\Omega,
\\
& -\omega B_\Pi = -k^2 \Pi_0 A_\Theta - \frac{k^2}{2} \Pi_0 \cos \Omega_0 A_\theta,\\
&  \omega A_\Pi = -k^2 \Pi_0 B_\Theta - \frac{k^2}{2} \Pi_0 \cos \Omega_0 B_\theta ,
\\
&-\omega B_\Omega = \left( \frac{k^2}{2} \sin \Omega_0 + 2 \delta  \right) A_\theta,  \\
&\omega A_\Omega = \left( \frac{k^2}{2} \sin \Omega_0 + 2 \delta  \right) B_\theta.
\end{cases}
\end{align*}

To find nontrivial solutions, the linear homogeneous system can be written compactly in matrix form:
\[
M \cdot 
\begin{pmatrix}
A_\Theta\\
B_\Theta  \\
A_\theta  \\
B_\theta\\
A_\Pi\\
B_\Pi \\
A_\Omega \\
B_\Omega
\end{pmatrix}
= 0,
\]
where the coefficient matrix \( M \) is:
\[\scalebox{0.80}{$
\begin{pmatrix}
0 & \omega & 0 & 0 & \frac{\alpha_1 + \alpha_2}{2} + \frac{k^2}{4\Pi_0} & 0 & \frac{k^2}{4} \cot\Omega_0 + \frac{\delta}{\sin\Omega_0} \cot\Omega_0 & 0 \\
-\omega & 0 & 0 & 0 & 0 & \frac{\alpha_1 + \alpha_2}{2} + \frac{k^2}{4\Pi_0} & 0 & \frac{k^2}{4} \cot\Omega_0 + \frac{\delta}{\sin\Omega_0} \cot\Omega_0 \\
0 & 0 & 0 & \omega &  (\alpha_1 - \alpha_2) \cos\Omega_0 & 0 & -(\alpha_1 - \alpha_2) \Pi_0 \sin\Omega_0 - \frac{k^2}{2 \sin\Omega_0} -\frac{2 \delta}{\sin^2\Omega_0} & 0 \\
0 & 0 & -\omega & 0 & 0 &  (\alpha_1 - \alpha_2) \cos\Omega_0 & 0 & -(\alpha_1 - \alpha_2) \Pi_0 \sin\Omega_0 - \frac{k^2}{2 \sin\Omega_0} -  \frac{2\delta}{\sin^2\Omega_0} \\
-k^2 \Pi_0 & 0 & -\frac{k^2}{2} \Pi_0 \cos\Omega_0 & 0 & 0 & \omega & 0 & 0 \\
0 & -k^2 \Pi_0 & 0 & -\frac{k^2}{2} \Pi_0 \cos\Omega_0 & -\omega & 0 & 0 & 0 \\
0 & 0 &  \frac{k^2}{2} \sin\Omega_0 + 2 \delta   & 0 & 0 & 0 & 0 & \omega \\
0 & 0 & 0 &  \frac{k^2}{2} \sin\Omega_0 + 2 \delta   & 0 & 0 & -\omega & 0 \\
\end{pmatrix}. $}
\]

Nontrivial solutions exist only when
\[
\det M(k, \omega) = 0,
\]
which determines the dispersion relation \( \omega(k) \) for elementary excitations.

Before considering the general scenario with $\delta \neq 0$ and $\Delta \neq 0$, we will examine particular cases to gain analytical insight into the spectrum.
\subsection*{3.1. Case $\delta=0$, $\Delta\neq0$, no relaxation}

We begin by considering the case $\delta=0$ and $\Delta \neq 0$ corresponding to a symmetric cavity placed in the external magnetic field along the z axis. In this case, the chemical potential and the condensate phase angle are given by: 
\begin{align} \label{conditions}
\mu= \frac{\alpha_1+\alpha_2}{2}{\Pi_0} \quad  \quad \quad \cos \Omega_0=\frac{\Delta}{(\alpha_1-\alpha_2)\Pi_0}.
\end{align}

The corresponding Bogoliubov matrix $M$ takes the form:
\[\scalebox{0.9}{$
\begin{pmatrix}
0 & \omega & 0 & 0 & \frac{\alpha_1 + \alpha_2}{2} + \frac{k^2}{4\Pi_0} & 0 & \frac{k^2}{4} \cot\Omega_0  & 0 \\
-\omega & 0 & 0 & 0 & 0 & \frac{\alpha_1 + \alpha_2}{2} + \frac{k^2}{4\Pi_0} & 0 & \frac{k^2}{4} \cot\Omega_0  \\
0 & 0 & 0 & \omega & (\alpha_1 - \alpha_2) \cos\Omega_0 & 0 & -(\alpha_1 - \alpha_2) \Pi_0 \sin\Omega_0 - \frac{k^2}{2 \sin\Omega_0}  & 0 \\
0 & 0 & -\omega & 0 & 0 & (\alpha_1 - \alpha_2) \cos\Omega_0 & 0 & -(\alpha_1 - \alpha_2) \Pi_0 \sin\Omega_0 - \frac{k^2}{2 \sin\Omega_0} \\
-k^2 \Pi_0 & 0 & -\frac{k^2}{2} \Pi_0 \cos\Omega_0 & 0 & 0 & \omega & 0 & 0 \\
0 & -k^2 \Pi_0 & 0 & -\frac{k^2}{2} \Pi_0 \cos\Omega_0 & -\omega & 0 & 0 & 0 \\
0 & 0 &  \frac{k^2}{2} \sin\Omega_0    & 0 & 0 & 0 & 0 & \omega \\
0 & 0 & 0 &  \frac{k^2}{2} \sin\Omega_0    & 0 & 0 & -\omega & 0 \\
\end{pmatrix} $}.
\]

Substituting the expressions \ref{conditions} for $\mu$ and $\cos \Omega_0$, and introducing the characteristic spin-splitting frequency $\omega_c = \frac{\alpha_1 - \alpha_2}{2} \Pi_0$, we apply an elementary row transformation to bring the matrix into a simplified form:
\[\scalebox{0.9}{$
M =
\begin{pmatrix}
-\omega & 0 & 0 & 0 & 0 & \frac{\mu}{\Pi_0} + \frac{k^2}{4\Pi_0} & 0 & \frac{k^2}{4} \cot\Omega_0  \\
0 & \omega & 0 & 0 & \frac{\mu}{\Pi_0} + \frac{k^2}{4\Pi_0} & 0 & \frac{k^2}{4} \cot\Omega_0  & 0 \\
0 & 0 & -\omega & 0 & 0 &  \frac{\Delta}{\Pi_0} & 0 & -2 \omega_c \sin\Omega_0 - \frac{k^2}{2 \sin\Omega_0} \\
0 & 0 & 0 & \omega &  \frac{\Delta}{\Pi_0} & 0 & -2 \omega_c \sin\Omega_0 - \frac{k^2}{2 \sin\Omega_0}  & 0 \\
0 & -k^2 \Pi_0 & 0 & -\frac{k^2 \Delta \Pi_0}{4 \omega_c} & -\omega & 0 & 0 & 0 \\
-k^2 \Pi_0 & 0 & -\frac{k^2 \Delta \Pi_0}{4 \omega_c} & 0 & 0 & \omega & 0 & 0 \\
0 & 0 & 0 &  \frac{k^2}{2} \sin\Omega_0    & 0 & 0 & -\omega & 0 \\
0 & 0 &  \frac{k^2}{2} \sin\Omega_0    & 0 & 0 & 0 & 0 & \omega \\
\end{pmatrix} $}.
\]

To evaluate the determinant of $M$, we rewrite it in block matrix form:
\[
M =
\begin{pmatrix}
W & A \\
B & W \\
\end{pmatrix}, 
\]
where $W, \ A, \ B$ are 4x4 matrix blocks. Hence, calculating the determinant of 8x8 matrix M is equivalent to calculating the determinant of 4x4 matrix, thanks to following formula
$$\det M= \det \begin{pmatrix}
W & A \\
B & W \\
\end{pmatrix} = \det W \det (W-BW^{-1}A).$$ Here, $W$ is diagonal with $\det W = \omega^4$, and the matrices $A$ and $B$ are:
\[\scalebox{1.1}{$
A =
\begin{pmatrix}
0 & d & 0 & f  \\
d & 0 & f & 0  \\
0 & e & 0 & \tilde{k}  \\
e & 0 & \tilde{k} & 0  \\
\end{pmatrix}  \quad \text{and} \quad 
B =
\begin{pmatrix}
0 & a& 0 & b  \\
a & 0 & b & 0  \\
0 & 0 & 0 & c  \\
0 & 0 & c & 0  \\
\end{pmatrix}, $}
\]
with the coefficients:$$\begin{cases} a=-k^2 \Pi_0,\\
b=-\frac{k^2 \Delta \Pi_0}{4 \omega_c}, \\
c=\frac{k^2}{2} \sin \Omega_0, \\
d = \frac{\mu}{\Pi_0}+\frac{k^2 }{4 \Pi_0},\\
f=\frac{k^2}{4} \cot \Omega_0, \\
e =  \frac{\Delta}{\Pi_0},\\
\tilde{k}=-2 \omega_c \sin \Omega_0 - \frac{k^2}{2 \sin \Omega_0}.
\end{cases}$$

It is not difficult to calculate the values of the matrix $W-BW^{-1}A$:
\[\scalebox{1.1}{$
W-BW^{-1}A =
\begin{pmatrix}
-(\omega + \frac{a}{\omega}d+\frac{b}{\omega}e) & 0 & -(\frac{a}{\omega}f+\frac{b}{\omega}\tilde{k}) & 0  \\
0 &  \omega + \frac{a}{\omega}d+\frac{b}{\omega}e  &0 & \frac{a}{\omega}f+\frac{b}{\omega}\tilde{k} \\
-\frac{c}{\omega}e & 0 & -(\omega+\frac{c}{\omega}\tilde{k}) & 0 \\
0 & \frac{c}{\omega}e   & 0 & \omega+\frac{c}{\omega}\tilde{k}. \\
\end{pmatrix}$}
\]

Consequently,
\begin{align} 
    &\det W \det (W-BW^{-1}A)=\omega^4 \left[ (\omega + \frac{a}{\omega}d+\frac{b}{\omega}e)^2(\omega+\frac{c}{\omega}\tilde{k})^2-2(\omega + \frac{a}{\omega}d+\frac{b}{\omega}e) (\omega+\frac{c}{\omega}\tilde{k}) (\frac{a}{\omega}f+\frac{b}{\omega}\tilde{k})(\frac{c}{\omega}e)+ (\frac{a}{\omega}f+\frac{b}{\omega}\tilde{k})^2(\frac{c}{\omega}e)^2 \right]= \notag \\ 
    &=  \left( (\omega^2 + ad+be)(\omega^2+c\tilde{k})-(af+b \tilde{k})ce \right)^2=0 \Rightarrow (\omega^2 + ad+be)(\omega^2+c\tilde{k})=(af+b \tilde{k})ce \notag \\
    &\Rightarrow \omega^4 + \omega^2( ad+be+c\tilde{k})+ ac(d\tilde{k}-fe)=0 \Rightarrow \omega^2 = - \frac{ ad+be+c\tilde{k}}{2} \pm \sqrt{\frac{( ad+be+c\tilde{k})^2}{4}-ac(d \tilde{k}-fe)} \label{omega}
\end{align}

By substituting the expressions for \(a, b, c, d, e, f, \tilde{k}\), we obtain the explicit form of the excitation spectrum:
\begin{align} 
    &\omega^2=\omega_0^2+\Pi_0 \left[ \alpha_1 \pm \sqrt{\alpha_2^2+\frac{\alpha_1^2-\alpha_2^2}{4 \omega_c^2}\Delta^2}\right]\omega_0,
\end{align}
where $\omega_0 = \frac{k^2}{2}$ is the bare polariton dispersion with $\omega_0(0) = 0$..
\subsection*{3.2. Case $\delta\neq0$, $\Delta=0$, no relaxation}  
In this section, we examine the regime characterized by finite in-plane spin anisotropy $\delta \neq 0$, vanishing Zeeman splitting $\Delta = 0$, and no relaxation. This corresponds to a typical experimental situation where the system exhibits anisotropic spin interactions in the absence of an external magnetic field. In this case, the stationary value of $\Omega_0$ is determined from:
$$ (\alpha_1-\alpha_2)\Pi_0 \cos \Omega_0+2\delta \cot \Omega_0   =0 \Rightarrow \cos \Omega_0 \left( (\alpha_1-\alpha_2)\Pi_0  +\frac{2\delta}{\sin \Omega_0}  \right)= 0 \Rightarrow \cos \Omega_0=0 \Rightarrow \Omega_0=\frac{\pi}{2}.$$ 

The expression for chemical potential $\mu$ can be written as: $$\mu= \frac{\alpha_1+\alpha_2}{2}{\Pi_0}-\delta. $$

The matrix \( M \) takes the form:
\[\scalebox{1.0}{$
\begin{pmatrix}
0 & \omega & 0 & 0 & \frac{\alpha_1 + \alpha_2}{2} + \frac{k^2}{4\Pi_0} & 0 & 0 & 0 \\
-\omega & 0 & 0 & 0 & 0 & \frac{\alpha_1 + \alpha_2}{2} + \frac{k^2}{4\Pi_0} & 0 & 0\\
0 & 0 & 0 & \omega &  0 & 0 & -(\alpha_1 - \alpha_2) \Pi_0 - \frac{k^2}{2 } -2 \delta & 0 \\
0 & 0 & -\omega & 0 & 0 &  0 & 0 & -(\alpha_1 - \alpha_2) \Pi_0  - \frac{k^2}{2 } -  2\delta \\
-k^2 \Pi_0 & 0 & 0 & 0 & 0 & \omega & 0 & 0 \\
0 & -k^2 \Pi_0 & 0 & 0 & -\omega & 0 & 0 & 0 \\
0 & 0 &  \frac{k^2}{2}  + 2 \delta   & 0 & 0 & 0 & 0 & \omega \\
0 & 0 & 0 &  \frac{k^2}{2}  + 2 \delta   & 0 & 0 & -\omega & 0\\
\end{pmatrix}. $}
\]

Identifying coefficients:
$$\begin{cases} a=-k^2 \Pi_0,\\
b=0, \\
c=\frac{k^2}{2} +2\delta, \\
d = \frac{\alpha_1+\alpha_2}{2}+\frac{k^2 }{4 \Pi_0},\\
f=0, \\
e =  0,\\
\tilde{k}=-(\alpha_1- \alpha_2) \Pi_0  - \frac{k^2}{2 } -  2\delta.
\end{cases}$$

Then the dispersion of elementary excitations will take the form
\begin{align}
\omega^2 = \omega_0^2 + \omega_0(\Pi_0 \alpha_1 + 2\delta) + 2\delta \omega_c + 2\delta^2 
\pm \left[ \omega_0(\Pi_0 \alpha_2 - 2\delta) - 2\delta(\omega_c + \delta) \right].
\end{align}

Thus, the two branches of the spectrum are:
\[
\begin{cases}
\omega_+^2 = \omega_0^2 + \omega_0 \Pi_0 (\alpha_1 + \alpha_2), \\
\omega_-^2 = (\omega_0 + 2\delta)^2 + 2\omega_c(\omega_0 + 2\delta).
\end{cases}
\]

The value of the gap between two dispersion branches at $k=0$ is concentration dependent and is given by the following formula:
\begin{align}
E_g=  \sqrt{4\delta \omega_c +4\delta^2}.
\end{align}
\subsection*{3.3. Case $\delta\neq0$, $\Delta\neq0$, no relaxation}
In this section, we analyze the more general case where both the in-plane spin anisotropy $\delta \neq 0$ and the Zeeman splitting $\Delta \neq 0$ are present, in the absence of relaxation. The hydrodynamic equations become:
\begin{align} 
\begin{cases}    \label{finalsystemwithoutrelax}
&\partial_t \Theta= \frac{1}{2} \left[ (\nabla \Theta)^2 +  \left( \frac{\nabla \theta}{2} \right)^2 \right] - \frac{1}{8} \left[ (\frac{\nabla \Pi}{\Pi}+ \nabla \Omega \cot{\Omega})^2 +\frac{(\nabla \Omega)^2}{\sin^2 \Omega} \right] + \frac{\alpha_1+\alpha_2}{2}{\Pi} - \frac{1}{4} \nabla  \left[ \frac{\nabla \Pi}{\Pi}+\nabla \Omega \cot{\Omega} \right]  -\frac{\delta \cos \theta }{\sin \Omega}  ,\\ \\
&\partial_t \theta= \nabla \Theta \nabla \theta +(\alpha_1-\alpha_2)\Pi \cos \Omega-\Delta + \frac{1}{2} \frac{\nabla \Omega}{\sin \Omega} \left[ \frac{\nabla \Pi }{\Pi}+\nabla \Omega \cot{\Omega}  \right]  +\frac{1}{2} \nabla  (\frac{\nabla \Omega}{\sin \Omega}) + 2 \delta \cot \Omega \cos \theta, \\
\\
&\partial_t \Pi=\nabla(\Pi\nabla \Theta)+\frac{1}{2}\nabla(\Pi \cos \Omega \nabla \theta),\\
\\
&\partial_t \Omega = \nabla \Omega  \nabla \Theta - \frac{\nabla  \Pi \nabla \theta}{2 \Pi} \sin \Omega -\frac{\nabla^2 \theta}{2} \sin \Omega-\frac{1}{2} \cos \Omega \nabla \Omega \nabla \theta +2 \delta \sin \theta.
\end{cases}
\end{align}

In the stationary regime:
\begin{align*}
\begin{cases}   
&\partial_t \Theta_0= \frac{\alpha_1+\alpha_2}{2}{\Pi_0}-\frac{\delta \cos \theta_0}{\sin \Omega_0}=\mu,\\ \\
&\partial_t \theta_0=  (\alpha_1-\alpha_2)\Pi_0 \cos \Omega_0-\Delta+2\delta \cot \Omega_0 \cos \theta_0  =0 ,\\
\\
&\partial_t \Pi_0=0,\\
\\
&\partial_t \Omega_0 =   2\delta \sin \theta_0=0.
\end{cases}
\end{align*}

From the last equation we find the restriction $\sin \theta_0 = 0 \Rightarrow \theta_0 = 0$. Substituting this into the second equation gives the condition:
\begin{align} \label{cond1}
(\alpha_1 - \alpha_2)\Pi_0 \cos \Omega_0 - \Delta + 2\delta \cot \Omega_0 = 0.
\end{align}
And the chemical potential simplifies to:
\begin{align}  \label{cond2}
\mu = \frac{\alpha_1 + \alpha_2}{2} \Pi_0 - \frac{\delta}{\sin \Omega_0}
\end{align}
By following a procedure similar to that described in the previous sections, we arrive at a matrix \( M \)  of the following form
\[\scalebox{0.8}{$
\begin{pmatrix}
-\omega & 0 & 0 & 0 & 0 & \frac{\alpha_1 + \alpha_2}{2} + \frac{k^2}{4\Pi_0} & 0 & \frac{k^2}{4} \cot\Omega_0 + \frac{\delta}{\sin\Omega_0} \cot\Omega_0 \\
0 & \omega & 0 & 0 & \frac{\alpha_1 + \alpha_2}{2} + \frac{k^2}{4\Pi_0} & 0 & \frac{k^2}{4} \cot\Omega_0 + \frac{\delta}{\sin\Omega_0} \cot\Omega_0 & 0  \\
 0 & 0 & -\omega & 0 & 0 &  (\alpha_1 - \alpha_2) \cos\Omega_0 & 0 & -(\alpha_1 - \alpha_2) \Pi_0 \sin\Omega_0 - \frac{k^2}{2 \sin\Omega_0} -  \frac{2\delta}{\sin^2\Omega_0}\\
0 & 0 & 0 & \omega &  (\alpha_1 - \alpha_2) \cos\Omega_0 & 0 & -(\alpha_1 - \alpha_2) \Pi_0 \sin\Omega_0 - \frac{k^2}{2 \sin\Omega_0} -\frac{2 \delta}{\sin^2\Omega_0} & 0  \\
0 & -k^2 \Pi_0 & 0 & -\frac{k^2}{2} \Pi_0 \cos\Omega_0 & -\omega & 0 & 0 & 0\\
-k^2 \Pi_0 & 0 & -\frac{k^2}{2} \Pi_0 \cos\Omega_0 & 0 & 0 & \omega & 0 & 0   \\
0 & 0 & 0 &  \frac{k^2}{2} \sin\Omega_0 +2 \delta   & 0 & 0 & -\omega & 0 \\
0 & 0 &  \frac{k^2}{2} \sin\Omega_0 + 2 \delta   & 0 & 0 & 0 & 0 & \omega   \\
\end{pmatrix} $}
\]

To find the spectrum of elementary excitations, we use the general dispersion relation derived earlier (see formula \ref{omega}), but we must redefine the coefficients \( a, b, c, d, \ldots \), which now take the following forms:
\begin{equation*}
    \begin{cases} 
    a=-k^2 \Pi_0,\\
    b=-\frac{k^2}{2} \Pi_0 \cos\Omega_0 , \\
    c=\frac{k^2}{2} \sin \Omega_0+2 \delta, \\
    d = \frac{\alpha_1+\alpha_2}{2}+\frac{k^2 }{4 \Pi_0},\\
    f=\cot \Omega_0(\frac{k^2}{4}+\frac{\delta}{\sin \Omega_0}) , \\
    e = (\alpha_1-\alpha_2)\cos \Omega_0,\\
    \tilde{k}=-2 \omega_c \sin\Omega_0 - \frac{k^2}{2 \sin\Omega_0} -\frac{2 \delta}{\sin^2\Omega_0}.
\end{cases}
\end{equation*}

Finally, the dispersion of elementary excitations in the case under consideration is written as
\begin{align}
\omega^2_{\pm} &= \omega_0^2 + \omega_0 \left[\Pi_0 \alpha_1 + \frac{2 \delta}{\sin \Omega_0} \right] 
+ 2 \delta \omega_c \sin \Omega_0 + \frac{2 \delta^2}{\sin^2 \Omega_0} \pm \\ \notag
&\quad \pm \bigg\{ 
\omega_0^2 \bigg[
\Pi_0^2\left(\alpha_1^2 \cos^2 \Omega_0 + \alpha_2^2 \sin^2 \Omega_0\right) 
- \frac{4 \delta \Pi_0}{\sin \Omega_0}
 (\alpha_1 \cos^2 \Omega_0 + \alpha_2 \sin^2 \Omega_0)
\bigg] - \\ \notag
&\quad - 4 \delta \omega_0 \Pi_0 \bigg[
\omega_c \alpha_2 \sin \Omega_0  
+ \delta  \left(\alpha_1 \cot^2 \Omega_0 + \alpha_2 \right)
\bigg]  + 4 \delta^2 \left(\frac{\omega_0}{\sin \Omega_0}+\omega_c \sin \Omega_0 + \frac{\delta}{\sin^2 \Omega_0} \right)^2
\bigg\}^{1/2}. \notag
\end{align}
where $\omega_c=\frac{\alpha_1-\alpha_2}{2}\Pi_0$ as before and $\Omega_0$ satisfies the condition: $2 \omega_c \cos \Omega_0-\Delta+2\delta \cot \Omega_0   =0$ .

The value of the gap between two dispersion branches at $k=0$ depends on the concentration and the external magnetic field. It is given by the following formula:
\begin{align} \label{E_g}
E_g=  \sqrt{4\delta \omega_c \sin \Omega_0+\frac{4\delta^2}{\sin^2\Omega_0}}.
\end{align}
\subsection*{3.4. Case $\delta\neq0$, $\Delta\neq0$ and non-zero relaxation terms} 
Now we proceed to compute the dispersion relation of  elementary excitations in the presence of relaxation effects. 
The full set of equations of motion, which includes relaxation and is valid for  $\delta \ne 0$ and $\Delta \ne 0$, was derived earlier (see Eq.\ref{systemwithrelax0}). It reads:
\begin{align} \label{systemwithrelax}
\begin{cases}    
&\partial_t \Theta= \frac{1}{2} \left[ (\nabla \Theta)^2 +  \left( \frac{\nabla \theta}{2} \right)^2 \right] - \frac{1}{8} \left[ (\frac{\nabla \Pi}{\Pi}+ \nabla \Omega \cot{\Omega})^2 +\frac{(\nabla \Omega)^2}{\sin^2 \Omega} \right] + \\
&\qquad\qquad\qquad\qquad\qquad\qquad\qquad + \frac{\alpha_1+\alpha_2}{2}{\Pi} - \frac{1}{4} \nabla  \left[ \frac{\nabla \Pi}{\Pi}+\nabla \Omega \cot{\Omega} \right]  -\frac{\delta \cos \theta }{\sin \Omega} +\gamma_\Theta[\nabla(\Pi\nabla \Theta)+\frac{1}{2}\nabla(\Pi \cos \Omega \nabla \theta)]  ,\\ \\
&\partial_t \theta= \nabla \Theta \nabla \theta +(\alpha_1-\alpha_2)\Pi \cos \Omega-\Delta + \frac{1}{2} \frac{\nabla \Omega}{\sin \Omega} \left[ \frac{\nabla \Pi }{\Pi}+\nabla \Omega \cot{\Omega}  \right]  + \\
&\qquad\qquad\qquad\qquad\qquad\qquad\qquad\qquad \qquad\qquad\qquad +\frac{1}{2} \nabla  (\frac{\nabla \Omega}{\sin \Omega}) + 2 \delta \cot \Omega \cos \theta+ \gamma_\theta [\frac{1}{4}\nabla(\Pi\nabla \theta)+\frac{1}{2}\nabla(\Pi \cos \Omega \nabla \Theta)-\delta \Pi \sin \Omega \sin \theta] , \\
\\
&\partial_t \Pi=\nabla(\Pi\nabla \Theta)+\frac{1}{2}\nabla(\Pi \cos \Omega \nabla \theta),\\
\\
&\partial_t \Omega = \nabla \Omega  \nabla \Theta - \frac{\nabla  \Pi \nabla \theta}{2 \Pi} \sin \Omega -\frac{\nabla^2 \theta}{2} \sin \Omega-\frac{1}{2} \cos \Omega \nabla \Omega \nabla \theta +2 \delta \sin \theta + \\
&\quad \quad \quad \quad \quad \quad \quad  \quad \quad \quad \quad \quad \quad \quad  +\gamma_\Omega[\ \frac{1}{2} \Pi \sin\Omega\, \nabla \Theta  \nabla \theta
+ \frac{1}{2} (\alpha_1 - \alpha_2) \Pi^2 \cos\Omega \sin\Omega
- \frac{1}{2} \Pi \Delta \sin\Omega + \delta \Pi \cos \Omega \cos \theta
+ \frac{1}{4}\nabla (\Pi \nabla \Omega) ].
\end{cases}
\end{align}

In the case of a spatially homogeneous condensate with zero moment, corresponding to the stationary state, we can obtain:
\begin{align}
\begin{cases}   \label{system} 
&\partial_t \Theta_0=\frac{\alpha_1+\alpha_2}{2}{\Pi_0}-\frac{\delta \cos \theta_0}{\sin \Omega_0}=\mu,\\ \\
&\partial_t \theta_0=  (\alpha_1-\alpha_2)\Pi_0 \cos \Omega_0-\Delta+2\delta \cot \Omega_0 \cos \theta_0 -\gamma_\theta \delta \Pi_0 \sin \Omega_0 \sin \theta_0=0  ,\\
\\
&\partial_t \Pi_0=0,\\
\\
&\partial_t \Omega =2\delta \sin \theta_0+\gamma_\Omega [
\frac{1}{2} (\alpha_1 - \alpha_2) \Pi_0^2 \cos\Omega_0 \sin\Omega_0
- \frac{1}{2} \Pi_0 \Delta \sin\Omega_0+\delta \Pi_0 \cos \Omega_0 \cos \theta_0
]=0.
\end{cases}
\end{align}

Using the second equation from \ref{system} and substituting it into the fourth expression, we derive the following condition for a stationary state: $$2 \delta \sin \theta_0+\frac{\gamma_\Omega\gamma_\theta}{2}\Pi_0^2 \delta \sin^2 \Omega_0 \sin \theta_0=0 \Rightarrow \sin \theta_0=0 \Rightarrow \theta_0=0. $$

The chemical potential in this stationary state becomes: $$\mu= \frac{\alpha_1+\alpha_2}{2}{\Pi_0}-\frac{\delta }{\sin \Omega_0},$$
and the equation for the polarization angle $\Omega_0$ takes the form:
$$(\alpha_1-\alpha_2)\Pi_0 \cos \Omega_0-\Delta+2\delta \cot \Omega_0 =0.$$

Notably, these conditions exactly coincide with the stationary solutions without relaxation \ref{cond1}, \ref{cond2}, which is expected. Relaxation does not affect the ground state — only the dynamics of perturbations around it.
To study the dispersion of excitations, we introduce small perturbations around the stationary solution:
\begin{equation}
\begin{cases} \label{ansatz} 
    \Pi=\Pi_0+\updelta\Pi(\bm{r},t), \\
    \Theta=\mu t+\updelta\Theta(\bm{r},t), \\
    \theta=\theta_0+\updelta\theta(\bm{r},t), \\
    \Omega=\Omega_0+\updelta\Omega(\bm{r},t),
\end{cases}
\end{equation}
where $|\delta f| \ll |f_0|$ for $f=\Pi, \Theta, \theta \ \text{or} \ \Omega$. 
Substituting this ansatz \ref{ansatz} into the system \ref{systemwithrelax} and linearizing with respect to $\delta \Omega, \delta \Pi, \delta \Theta, \delta \theta$, we get
\begin{align*}
\begin{cases}   
&\partial_t \updelta \Theta=   \frac{\alpha_1+\alpha_2}{2}\updelta\Pi - \frac{1}{4} \frac{\nabla^2 \updelta \Pi}{\Pi_0}- \frac{1}{4} \cot{\Omega_0} \nabla^2 \updelta\Omega + \frac{\delta}{\sin \Omega_0}\cot\Omega_0 \updelta \Omega +\gamma_\Theta[\Pi_0\nabla^2\updelta\Theta+\frac{1}{2} \Pi_0 \cos \Omega_0 \nabla^2 \updelta \theta] ,\\ \\
&\partial_t \updelta \theta= - (\alpha_1-\alpha_2)[\Pi_0 \sin\Omega_0\updelta\Omega-\cos\Omega_0\updelta\Pi] +\frac{1}{2} \frac{\nabla^2 \updelta \Omega}{\sin \Omega_0} - \frac{2 \delta}{\sin^2\Omega_0}\updelta \Omega + \gamma_\theta [\frac{1}{4}\Pi_0 \nabla^2 \delta \theta+\frac{1}{2} \Pi_0 \cos \Omega_0 \nabla^2 \delta \Theta-\delta \Pi_0 \sin \Omega_0 \updelta \theta]  , \\
\\
&\partial_t \updelta\Pi= \Pi_0 \nabla^2 \updelta \Theta+\frac{1}{2} \Pi_0 \cos \Omega_0 \nabla^2 \updelta \theta,\\
\\
&\partial_t \updelta \Omega = -\frac{\nabla^2 \updelta \theta}{2} \sin \Omega_0+2 \delta \updelta \theta +\gamma_\Omega [
\frac{1}{2} (\alpha_1 - \alpha_2) (\Pi_0^2 \cos 2\Omega_0 \delta \Omega+\Pi_0 \delta \Pi \sin 2\Omega_0)
- \frac{1}{2} \Delta \Pi_0 \cos \Omega_0 \updelta \Omega - \frac{1}{2}\Delta \updelta \Pi \sin \Omega_0 
+ \frac{1}{4} \Pi_0 \nabla^2 \updelta \Omega].
\end{cases}
\end{align*}

Next, we represent the small deviations as a superposition of real harmonic functions, as introduced in Eq.~\eqref{harmonics}. Making use of the  identities provided in Eq.~\eqref{identities} and substituting this representation into the linearized system, we calculate the set of coupled equations governing the amplitudes of the perturbations, which is equivalent to a system of eight equations:
\begin{align*}
\begin{cases}
&-\omega B_\Theta = \frac{\alpha_1 + \alpha_2}{2} A_\Pi + \frac{k^2}{4 \Pi_0} A_\Pi + \frac{k^2}{4} \cot \Omega_0 A_\Omega +\frac{\delta}{\sin \Omega_0} \cot \Omega_0 A_\Omega -\gamma_\Theta k^2 \Pi_0 A_\Theta -\gamma_\Theta \frac{k^2}{2} \Pi_0 \cos \Omega_0 A_\theta, \\
&  \omega A_\Theta = \frac{\alpha_1 + \alpha_2}{2} B_\Pi + \frac{k^2}{4 \Pi_0} B_\Pi + \frac{k^2}{4} \cot \Omega_0 B_\Omega + \frac{\delta}{\sin \Omega_0}  \cot \Omega_0 B_\Omega-\gamma_\Theta k^2 \Pi_0 B_\Theta -\gamma_\Theta \frac{k^2}{2} \Pi_0 \cos \Omega_0 B_\theta, \\
&-\omega B_\theta = -(\alpha_1 - \alpha_2)\left( \Pi_0 \sin \Omega_0 A_\Omega - \cos \Omega_0 A_\Pi \right) - \frac{k^2}{2 \sin \Omega_0} A_\Omega - \frac{2 \delta}{\sin^2 \Omega_0} A_\Omega -\gamma_\theta \frac{k^2}{4} \Pi_0 A_\theta -\gamma_\theta \frac{k^2}{2}\Pi_0\cos \Omega_0 A_\Theta-\gamma_\theta \delta \Pi_0 \sin \Omega_0 A_\theta ,   \\
& 
\omega A_\theta = -(\alpha_1 - \alpha_2)\left( \Pi_0 \sin \Omega_0 B_\Omega - \cos \Omega_0 B_\Pi \right) - \frac{k^2}{2 \sin \Omega_0} B_\Omega -   \frac{2 \delta}{\sin^2 \Omega_0} B_\Omega -\gamma_\theta \frac{k^2}{4} \Pi_0 B_\theta -\gamma_\theta \frac{k^2}{2}\Pi_0\cos \Omega_0 B_\Theta-\gamma_\theta \delta \Pi_0 \sin \Omega_0 B_\theta,
\\
& -\omega B_\Pi = -k^2 \Pi_0 A_\Theta - \frac{k^2}{2} \Pi_0 \cos \Omega_0 A_\theta,\\
&  \omega A_\Pi = -k^2 \Pi_0 B_\Theta - \frac{k^2}{2} \Pi_0 \cos \Omega_0 B_\theta ,
\\
&-\omega B_\Omega = \left( \frac{k^2}{2} \sin \Omega_0 + 2 \delta  \right) A_\theta+\gamma_\Omega \frac{1}{2} (\alpha_1-\alpha_2)\Pi_0[\Pi_0 \cos 2\Omega_0 A_\Omega+\sin 2 \Omega_0 A_\Pi] -\frac{\gamma_\Omega}{2}\Delta \Pi_0 \cos \Omega_0 A_\Omega - \frac{\gamma_\Omega}{2} \Delta \sin \Omega_0 A_\Pi -\frac{k^2}{4} \gamma_\Omega \Pi_0 A_\Omega,  \\
&\omega A_\Omega = \left( \frac{k^2}{2} \sin \Omega_0 + 2 \delta  \right) B_\theta+\gamma_\Omega \frac{1}{2} (\alpha_1-\alpha_2)\Pi_0[\Pi_0 \cos 2\Omega_0 B_\Omega+\sin 2 \Omega_0 B_\Pi] -\frac{\gamma_\Omega}{2}\Delta \Pi_0 \cos \Omega_0 B_\Omega - \frac{\gamma_\Omega}{2} \Delta \sin \Omega_0 B_\Pi -\frac{k^2}{4} \gamma_\Omega \Pi_0 B_\Omega.
\end{cases}
\end{align*}

Let us associate this system of linear homogeneous equations with the matrix \( M \). For convenience, we perform a row permutation by swapping adjacent rows of the matrix and obtain
\[\scalebox{0.55}{$
\begin{pmatrix}
-\omega & -\gamma_\Theta k^2\Pi_0  & 0 & -\gamma_\Theta \frac{k^2}{2} \Pi_0 \cos \Omega_0 & 0 & \frac{\alpha_1 + \alpha_2}{2} + \frac{k^2}{4\Pi_0} & 0 & \frac{k^2}{4} \cot\Omega_0 + \frac{\delta}{\sin\Omega_0} \cot\Omega_0  \\
-\gamma_\Theta k^2\Pi_0 & \omega & -\gamma_\Theta \frac{k^2}{2} \Pi_0 \cos \Omega_0& 0 & \frac{\alpha_1 + \alpha_2}{2} + \frac{k^2}{4\Pi_0} & 0 & \frac{k^2}{4} \cot\Omega_0 + \frac{\delta}{\sin\Omega_0} \cot\Omega_0 & 0\\
0 & -\gamma_\theta \frac{k^2}{2} \Pi_0 \cos \Omega_0  & -\omega & -\gamma_\theta \Pi_0(\frac{k^2}{4}+\delta\sin \Omega_0)  & 0 &  (\alpha_1 - \alpha_2) \cos\Omega_0 & 0 & -(\alpha_1 - \alpha_2) \Pi_0 \sin\Omega_0 - \frac{k^2}{2 \sin\Omega_0} -  \frac{2\delta}{\sin^2\Omega_0} \\
-\gamma_\theta \frac{k^2}{2} \Pi_0 \cos \Omega_0 & 0 & -\gamma_\theta \Pi_0(\frac{k^2}{4}+\delta\sin \Omega_0) & \omega &  (\alpha_1 - \alpha_2) \cos\Omega_0 & 0 & -(\alpha_1 - \alpha_2) \Pi_0 \sin\Omega_0 - \frac{k^2}{2 \sin\Omega_0} -\frac{2 \delta}{\sin^2\Omega_0} & 0   \\
 0 & -k^2 \Pi_0 & 0 & -\frac{k^2}{2} \Pi_0 \cos\Omega_0 & -\omega & 0 & 0 & 0  \\
-k^2 \Pi_0 & 0 & -\frac{k^2}{2} \Pi_0 \cos\Omega_0 & 0 & 0 & \omega & 0 & 0 \\
0 & 0 & 0 &  \frac{k^2}{2} \sin\Omega_0 + 2 \delta   & 0 & \frac{\gamma_\Omega}{2}[ (\alpha_1-\alpha_2) \Pi_0 \sin 2 \Omega_0   - \Delta \sin \Omega_0 ] & -\omega & \frac{\gamma_\Omega}{2} \Pi_0[(\alpha_1-\alpha_2)\Pi_0 \cos 2\Omega_0  - \Delta \cos \Omega_0   -\frac{k^2}{2} ]  \\
0 & 0 &  \frac{k^2}{2} \sin\Omega_0 + 2 \delta   & 0 & \frac{\gamma_\Omega}{2}[ (\alpha_1-\alpha_2) \Pi_0 \sin 2 \Omega_0   - \Delta \sin \Omega_0 ]  & 0 & \frac{\gamma_\Omega}{2} \Pi_0[(\alpha_1-\alpha_2) \Pi_0 \cos 2\Omega_0  -\Delta \cos \Omega_0   -\frac{k^2}{2} ]  & \omega  \\
\end{pmatrix} $}
\]
The solution of the equation \( \det M(k, \omega) = 0 \) provides the dispersion relation of the elementary excitations, taking into account the relaxation terms.
\begin{figure}[h!]
\includegraphics[width=0.8\linewidth]{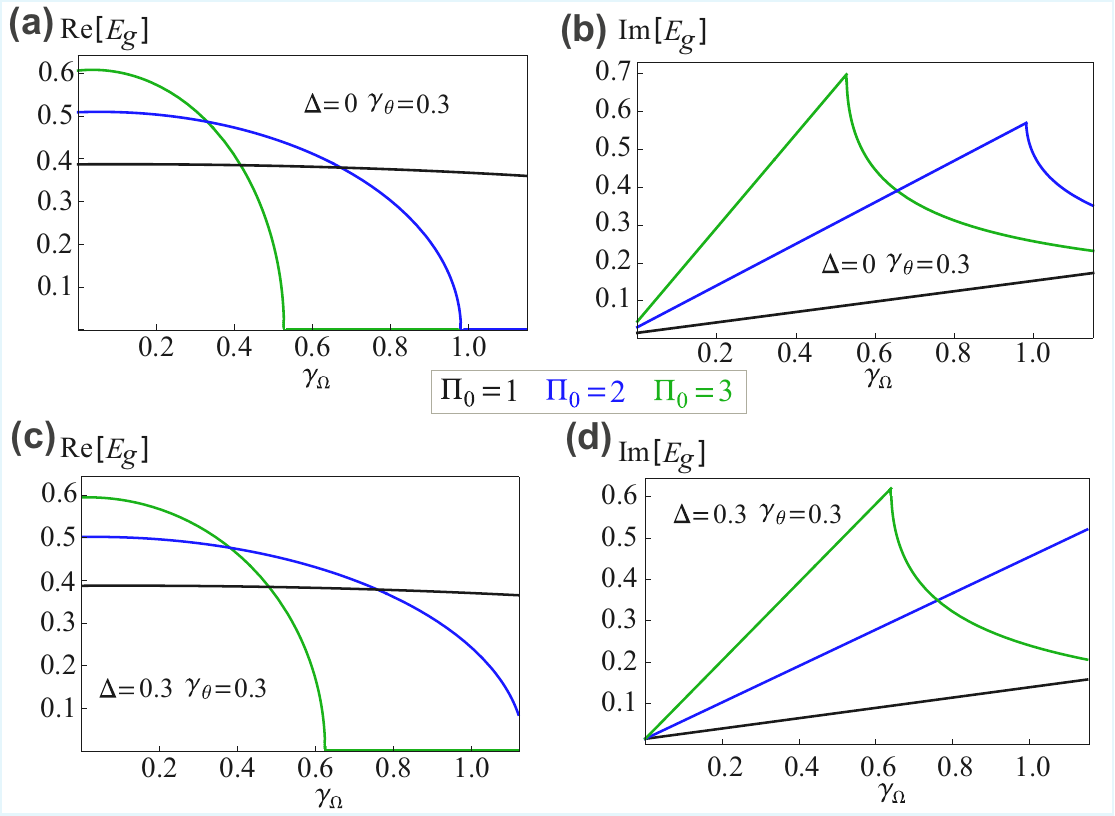}
\caption{Dependence of the real and imaginary parts of the excitation gap on the relaxation parameter \(\gamma_\Omega\) at fixed \(\gamma_\theta = 0.3\) for various densities \(\Pi_0\). The black curves correspond to \(\Pi_0 = 1\), while blue and green represent \(\Pi_0 = 2\) and \(\Pi_0 = 3\), respectively. Panels (a) and (b) illustrate the system's behavior in the absence of Zeeman splitting (\(\Delta = 0\)), while panels (c) and (d) correspond to the magnetically induced regime characterized by a finite Zeeman splitting \(\Delta = 0.3\). The other parameters are \(\alpha_1 = 0.5\), \(\alpha_2 = -0.1\alpha_1\), and \(\delta = 0.1\).  \label{gap1}}
\end{figure}

Let us analyze the behavior of the dispersion relation at the point \( k = 0 \). In this case, the matrix \( M \) takes the form:
\[\scalebox{0.67}{$
\begin{pmatrix}
-\omega & 0  & 0 & 0 & 0 & \frac{\alpha_1 + \alpha_2}{2}  & 0 & \frac{\delta}{\sin\Omega_0} \cot\Omega_0  \\
0 & \omega & 0 & 0 & \frac{\alpha_1 + \alpha_2}{2} & 0 &  \frac{\delta}{\sin\Omega_0} \cot\Omega_0 & 0\\
0 & 0 & -\omega & -\gamma_\theta \Pi_0 \delta\sin \Omega_0  & 0 &  (\alpha_1 - \alpha_2) \cos\Omega_0 & 0 & -(\alpha_1 - \alpha_2) \Pi_0 \sin\Omega_0 -\frac{2 \delta}{\sin^2\Omega_0} \\
0 & 0 & -\gamma_\theta \Pi_0\delta\sin \Omega_0 & \omega &  (\alpha_1 - \alpha_2) \cos\Omega_0 & 0 & -(\alpha_1 - \alpha_2) \Pi_0 \sin\Omega_0 -\frac{2 \delta}{\sin^2\Omega_0} & 0   \\
 0 & 0 & 0 & 0 & -\omega & 0 & 0 & 0  \\
0 & 0 & 0 & 0 & 0 & \omega & 0 & 0 \\
0 & 0 & 0 &   2 \delta   & 0 & \frac{\gamma_\Omega}{2}[ (\alpha_1-\alpha_2) \Pi_0 \sin 2 \Omega_0   - \Delta \sin \Omega_0 ] & -\omega & \frac{\gamma_\Omega}{2} \Pi_0[(\alpha_1-\alpha_2)\Pi_0 \cos 2\Omega_0  - \Delta \cos \Omega_0  ]  \\
0 & 0 &   2 \delta   & 0 & \frac{\gamma_\Omega}{2}[ (\alpha_1-\alpha_2) \Pi_0 \sin 2 \Omega_0   - \Delta \sin \Omega_0 ]  & 0 & \frac{\gamma_\Omega}{2} \Pi_0[(\alpha_1-\alpha_2) \Pi_0 \cos 2\Omega_0  -\Delta \cos \Omega_0  ]  & \omega  \\
\end{pmatrix}. $}
\]

We now compute the determinant of this matrix and solve the characteristic equation. The resulting dispersion relation has the form:
\begin{equation} \label{freqforzerok}
\omega_\pm^2 =
4\delta \omega_c \sin \Omega_0
+ \frac{4\delta^2}{\sin^2 \Omega_0}
- 2a^2 -2b^2
\pm 2(-a + b) \sqrt{(a + b)^2 - \left(4\delta \omega_c \sin \Omega_0 + \frac{4\delta^2}{\sin^2 \Omega_0}\right)},
\end{equation}
where 
\[
a = \frac{\gamma_\theta}{2} \Pi_0 \delta \sin \Omega_0, \qquad
b = \frac{\gamma_\Omega}{2} \Pi_0 \left( \omega_c \cos 2\Omega_0 - \frac{\Delta}{2} \cos \Omega_0 \right).
\]

In the absence of relaxation effects, i.e., for \( \gamma_\theta = 0 \) and \( \gamma_\Omega = 0 \), the expression simplifies considerably:
\[
\omega_+^2 = \omega_-^2 =
4\delta \omega_c \sin \Omega_0
+ \frac{4\delta^2}{\sin^2 \Omega_0},
\]
and the determinant becomes:
\[
\det M = \omega^4 (\omega^2 - \omega_+^2)^2 = 0.
\]
This yields a gap in the excitation spectrum:
\[
E_g = \sqrt{4\delta \omega_c \sin \Omega_0 + \frac{4\delta^2}{\sin^2 \Omega_0}},
\]

which is consistent with equation~\ref{E_g} obtained earlier.

Next, we will examine the evolution of the excitation gap as a function of key system parameters. At zero wave vector, the linearized spectrum consists of five distinct frequencies, as indicated in Eq.~\eqref{freqforzerok}: a fourfold-degenerate zero mode \(\omega = 0\), and two pairs of symmetric modes at \(\pm \omega_-\) and \(\pm \omega_+\). In what follows, we define the excitation gap as the value of \(\omega_-\), i.e., the distance between this lowest-frequency mode and zero.

Figure~\ref{gap1} presents the dependence of the real and imaginary parts of the gap on the relaxation coefficient \(\gamma_\Omega\), at a fixed value \(\gamma_\theta = 0.3\), for different polariton densities \(\Pi_0\). Panels (a) and (b) correspond to the case without Zeeman splitting (\(\Delta = 0\)). One observes that the critical value of \(\gamma_\Omega\) at which the real part of the gap vanishes decreases with increasing \(\Pi_0\). Notably, at the point of gap closure, the imaginary part of the spectrum undergoes a qualitative transition: its monotonic growth is replaced by a subsequent decrease, indicating a change in the nature of the elementary excitations.

Panels (c) and (d) of Fig.~\ref{gap1} depict the same dependencies in the presence of an external magnetic field (\(\Delta = 0.3\)). The inclusion of Zeeman splitting leads to an increase in the critical value of \(\gamma_\Omega\) required to close the gap, thereby stabilizing the spectrum against relaxation-induced softening.

\begin{figure}[h!]
\includegraphics[width=0.8\linewidth]{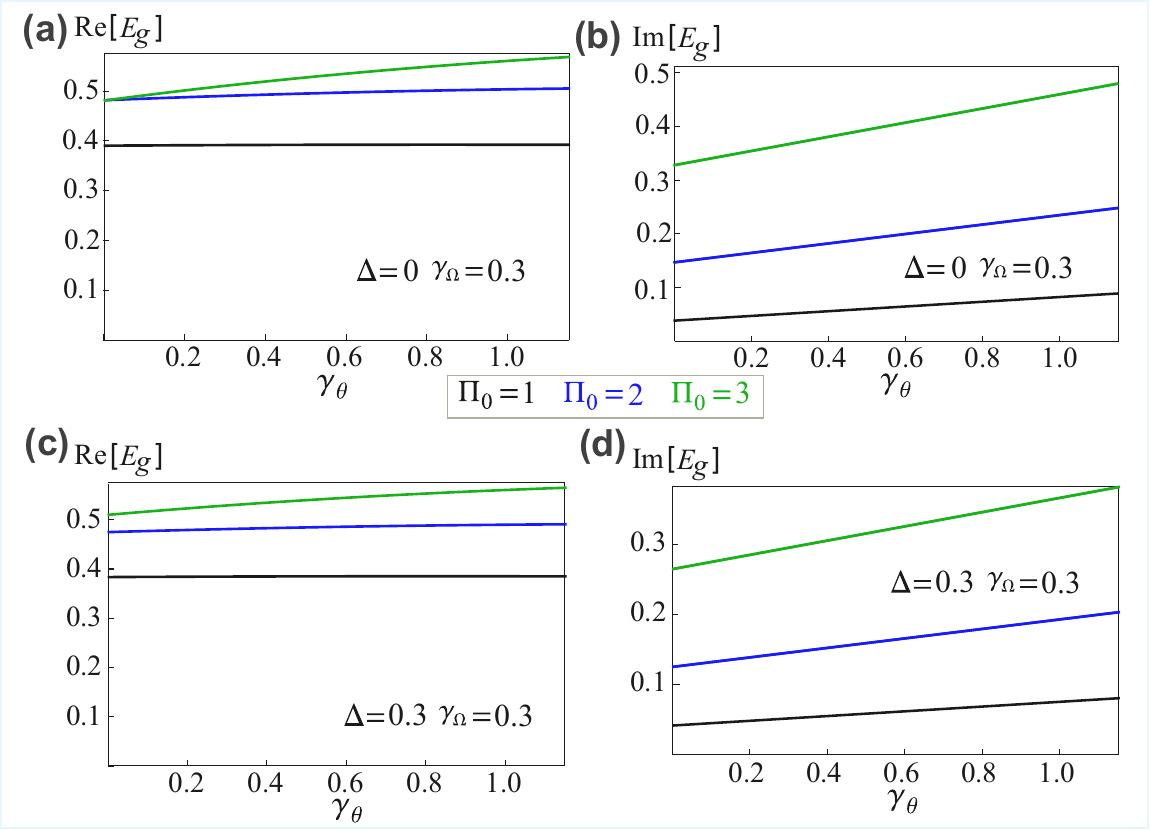}
\caption{Dependence of the real and imaginary parts of the excitation gap on the relaxation parameter \(\gamma_\theta\) at fixed \(\gamma_\Omega = 0.3\) for various densities \(\Pi_0\). The black, blue, and green curves correspond to \(\Pi_0 = 1\), \(2\), and \(3\), respectively. Panels (a) and (b) show the case of zero Zeeman splitting (\(\Delta = 0\)), whereas panels (c) and (d) illustrate the case of a microcavity subject to a magnetic field with Zeeman splitting \(\Delta = 0.3\). The remaining parameters are \(\alpha_1 = 0.5\), \(\alpha_2 = -0.1\alpha_1\), and \(\delta = 0.1\). \label{gap2}}
\end{figure}

A different picture emerges in Fig.~\ref{gap2}, which shows the gap's behavior as a function of \(\gamma_\theta\), at fixed \(\gamma_\Omega = 0.3\). For all densities considered, and both in the absence [panels (a), (b)] and presence [panels (c), (d)] of a magnetic field, the real part of the gap increases linearly with \(\gamma_\theta\), while the imaginary part exhibits a mild growth. Unlike the \(\gamma_\Omega\) case, no gap closure is observed within the parameter range studied, indicating that \(\gamma_\theta\) enhances spectral stability.

Together, these figures demonstrate how both relaxation mechanisms and external magnetic fields can significantly influence the excitation spectrum, altering not only the gap magnitude but also the onset of dynamical instabilities.